\documentclass[preprint2]{proto}
\usepackage{times}
\usepackage{natbib}

\begin{document}

\title{\textbf{\LARGE FORMATION AND COLLISIONAL EVOLUTION OF KUIPER BELT OBJECTS}}

\author {\textbf{\large Scott J. Kenyon}}
\affil{\small\em Smithsonian Astrophysical Observatory}
\author {\textbf{\large Benjamin C. Bromley}}
\affil{\small\em Department of Physics, University of Utah}
\author {\textbf{\large David P. O'Brien}}
\affil{\small\em Planetary Science Institute}
\author {\textbf{\large Donald R. Davis}}
\affil{\small\em Planetary Science Institute}

\begin{abstract}
\baselineskip = 11pt
\leftskip = 0.65in 
\rightskip = 0.65in
\parindent=1pc
{\small 
This chapter summarizes analytic theory and numerical calculations for the 
formation and collisional evolution of KBOs at 20--150 AU.  We describe the 
main predictions of a baseline self-stirring model and show how dynamical
perturbations from a stellar flyby or stirring by a giant planet modify the
evolution.  Although robust comparisons between observations and theory require
better KBO statistics and more comprehensive calculations, the data are broadly
consistent with KBO formation in a massive disk followed by substantial 
collisional grinding and dynamical ejection.  However, there are important 
problems reconciling the results of coagulation and dynamical calculations. 
Contrasting our current understanding of the evolution of KBOs and asteroids 
suggests that additional observational constraints, such as the identification 
of more dynamical families of KBOs (like the 2003 EL61 family), would provide 
additional information on the relative roles of collisional grinding and 
dynamical ejection in the Kuiper Belt.  The uncertainties also motivate 
calculations that combine collisional and dynamical evolution, a `unified' 
calculation that should give us a better picture of KBO formation and evolution.
\\~\\~\\~}%leave this in to get the correct vertical space after the abstract
\end{abstract}  

\section{\textbf{INTRODUCTION}}

Every year in the Galaxy, a star is born.  Most stars form in dense 
clusters of thousands of stars, as in the Orion Nebula Cluster 
\citep{lada2003,sles2004}.  Other stars form in small groups of 5--10 
stars in loose associations of hundreds of stars, as in the Taurus-Auriga 
clouds \citep{gomez1993,luh2006}.  Within these associations and clusters, 
most newly-formed massive stars are binaries; lower mass stars are usually 
single \citep{lada2006}.

Large, optically thick circumstellar disks surround nearly all 
newly-formed stars \citep{beck1996}.  The disks have sizes of 
$\sim$ 100--200 AU, masses of $\sim$ 0.01-0.1 $M_{\odot}$, and 
luminosities of $\sim$ 0.2--1 $L_{\star}$, where $L_{\star}$ is 
the luminosity of the central star. The masses and geometries of 
these disks are remarkably similar to the properties of the minimum
mass solar nebula (MMSN), the disk required for the planets in the 
solar system \citep{weid1977a,hay1981,scholz2006}.

As stars age, they lose their disks. For solar-type stars, radiation 
from the opaque disk disappears in 1--10 Myr \citep{haisch2001}.
Many older stars have optically thin {\it debris disks} comparable 
in size to the opaque disks of younger stars but with much smaller 
masses, $\lesssim ~ 1 ~ M_{\oplus}$, and luminosities, 
$\lesssim 10^{-3}$ $L_{\star}$ ({\em Chapter} by {\em Moro-Martin et al.}). 
The lifetime of this phase is uncertain. Some 100 Myr-old stars have
no obvious debris disk; a few 1--10 Gyr-old stars have massive debris 
disks \citep{greaves2005}.

In the current picture, planets form during the transition from an 
optically thick protostellar disk to an optically thin debris disk.
From the statistics of young stars in molecular clouds, the timescale 
for this transition, $\sim 10^5$ yr, is comparable to the timescales 
derived for the formation of planetesimals from dust grains 
\citep{weid1977b,youdin:2002,dulle2005} and for the formation of lunar-mass 
or larger planets from planetesimals \citep{weth1993,weid1997,kokubo:2000,
naga2005,kb2006}. Because the grains in debris disks have short collision lifetimes, 
$\lesssim$ 1 Myr, compared to the ages of their parent stars, $\gtrsim$ 
10 Myr, high velocity collisions between larger objects must maintain the 
small grain population \citep{aum1984,back1993}.  The inferred dust production 
rates for debris disks around 0.1--10 Gyr old stars, $\sim 10^{20}$ 
g yr$^{-1}$, require an initial mass in 1 km objects, $M_i \sim$ 10--100 
$M_{\oplus}$, comparable to the amount of solids in the MMSN.  Because 
significant long-term debris production also demands gravitational 
stirring by an ensemble of planets with radii of 500--1000 km or larger 
\citep{kb2004a,wyatt2005}, debris disks probably are newly-formed 
planetary systems \citep{aum1984, back1993,arty1997,kb2002b,kb2004a,kb2004b}.

KBOs provide a crucial test of this picture.  With objects ranging in size 
from 10--20 km to $\sim$ 1000 km, the size distribution of KBOs yields a 
key comparison with theoretical calculations of planet formation 
\citep{davis:kbo,kl1998,kl1999a,kl1999b}.  Once KBOs have sizes of 
100--1000 km, collisional grinding, dynamical perturbations by large 
planets and passing stars, and self-stirring by small embedded planets 
produce features in the distributions of sizes and dynamical elements 
that observations can probe in detail.  Although complete calculations 
of KBO formation and dynamical evolution are not available, these 
calculations will eventually yield a better understanding of planet 
formation at 20--100 AU.

The Kuiper belt also enables a vital link between the solar system and 
other planetary systems.  With an outer radius of $\gtrsim$ 1000 AU 
(Sedna's aphelion) and a current mass of $\sim$ 0.1 M$_{\oplus}$ 
\citep[][{\em Cahpter by Petit et al.}]{luu2002,bernstein:tnodist},
the Kuiper belt has properties similar to those derived for the oldest 
debris disks \citep{greav2004, wyatt2005}.  Understanding planet formation 
in the Kuiper belt thus informs our interpretation of evolutionary 
processes in other planetary systems.

This paper reviews applications of coagulation theory for planet 
formation in the Kuiper belt. After a brief introduction to the 
theoretical background in \S2, we describe results from numerical 
simulations in \S3, compare relevant KBO observations with the 
results of numerical simulations in \S4, and contrast the properties 
of KBOs and asteroids in \S5.  We conclude with a short summary in \S6.

\section{\textbf{COAGULATION THEORY}}

Planet formation begins with dust grains suspended in a gaseous 
circumstellar disk.  Grains evolve into planets in three steps.  
Collisions between grains produce larger aggregates which decouple 
from the gas and settle into a dense layer in the disk midplane.  
Continued growth of the loosely bound aggregates leads to 
planetesimals, gravitationally bound objects whose motions are 
relatively independent of the gas. Collisions and mergers among the 
ensemble of planetesimals form planets. Here, we briefly describe 
the physics of these stages and summarize analytic results as a 
prelude to summaries of numerical simulations.

We begin with a prescription for the mass surface density $\Sigma$ 
of gas and dust in the disk.  We use subscripts `d' for the dust and 
`g' for the gas and adopt
\begin{equation}
\Sigma_{\rm d,g} = \Sigma_{\rm 0d,0g} \left(\frac{a}{\rm 40 ~ AU}
\right)^{-n} ,
\label{eq:mmsn} \\
\end{equation}
where $a$ is the semimajor axis.  In the MMSN, $n$ = 3/2, 
$\Sigma_{\rm 0d}\approx 0.1$ $\rm g~cm^{-2}$, and 
$\Sigma_{\rm 0g} \approx $ 5--10 $\rm g~cm^{-2}$
\citep{weid1977a,hay1981}.
For a disk with an outer radius of 100 AU, the MMSN has a mass of 
$\sim$ 0.03 $M_{\odot}$, which is comparable to the disk masses of 
young stars in nearby star-forming regions \citep{natta2000,scholz2006}.

The dusty midplane forms quickly \citep{weid1977b,weid1980,dulle2005}.
For interstellar grains with radii, $r \sim$ 0.01--0.1 $\mu$m, turbulent 
mixing approximately balances settling due to the vertical component 
of the star's gravity. As grains collide and grow to $r \sim$ 0.1--1 mm, 
they decouple from the turbulence and settle into a thin layer in the 
disk midplane. The timescale for this process is $\sim 10^3$ yr at 1 AU 
and $\sim 10^5$ yr at 40 AU.

The evolution of grains in the midplane is uncertain. Because the gas 
has some pressure support, it orbits the star slightly more slowly than
the Keplerian velocity. Thus, orbiting dust grains feel a headwind that
drags them toward the central star \citep{ada76,weid1984,tanaka:1999}.  
For m-sized objects, the drag timescale at 40 AU, $\sim 10^5$ yr,
is comparable to the growth time. 
Thus, it is not clear whether grains can grow by direct accretion to 
km sizes before the gas drags them into the inner part of the disk.

Dynamical processes provide alternatives to random agglomeration of 
grains. 
In ensembles of porous grains, gas flow during disruptive collisions 
leads to planetesimal formation by direct accretion \citep{wurm2004}.
Analytic estimates and numerical simulations indicate that grains with 
$r \sim$ 1 cm are also easily trapped within vortices in the disk 
\citep[e.g.][]{dela2001,inaba2006}.  Large enhancements in the 
solid-to-gas ratio within vortices allows accretion to overcome gas 
drag, enabling formation of km-sized planetesimals in $10^4$--$10^5$ yr.

If the dusty midplane is calm, it becomes thinner and thinner until 
groups of particles overcome the local Jeans criterion -- where 
their self-gravity overcomes local orbital shear -- and `collapse' into 
larger objects on the local dynamical timescale, $\sim 10^3$ yr at 40 AU 
\citep{gold1973,youdin:2002,tanga2004}. This process is a promising way 
to form planetesimals; however, turbulence may prevent the instability 
\citep{weid1995,weid2003, weid2006}. Although the expected size of a 
collapsed object is the Jeans wavelength, the range of planetesimal 
sizes the instability produces is also uncertain.

Once planetesimals with $r \sim$ 1 km form, gravity dominates gas dynamics.
Long range gravitational interactions exchange kinetic energy (dynamical 
friction) and angular momentum (viscous stirring), redistributing orbital 
energy and angular momentum among planetesimals. For 1~km objects at 40 AU, 
the initial random velocities are comparable to their escape velocities, 
$\sim$ 1 m s$^{-1}$ \citep{weid1980,gold2004}.  The gravitational binding 
energy (for brevity, we use energy as a shorthand for specific energy), 
$E_g \sim 10^4$ erg g$^{-1}$, is then comparable to the typical 
collision energy, $E_c \sim 10^4$ erg g$^{-1}$. Both energies are 
smaller than the disruption energy -- the collision energy needed 
to remove half of the mass from the colliding pair of objects -- which 
is $Q_D^* \sim 10^5$--$10^7$ erg g$^{-1}$ for icy material \citep{davis:vesta,
benz1999,ryan1999,michel2001,lein2002,giblin2004}.
Thus, collisions produce mergers instead of debris. 

Initially, small planetesimals grow slowly. For a large ensemble of 
planetesimals, the collision rate is $n \sigma v$, where $n$ is the number 
of planetesimals, $\sigma$ is the cross-section, and $v$ is the relative 
velocity.  The collision cross-section is the geometric cross-section,
$\pi r^2$, scaled by the gravitational focusing factor, $f_g$,
\begin{equation}
\sigma_c \sim \pi r^2 f_g \sim \pi r^2 (1 + \beta (v_{esc}/e v_K)^2) ~ ,
\label{eq:cross}
\end{equation}
where $e$ is the orbital eccentricity, $v_K$ is the orbital velocity, 
$v_{esc}$ is the escape velocity of the merged pair of planetesimals, 
and $\beta \approx 2.7$ is a coefficient that accounts for three-dimensional 
orbits in a rotating disk \citep{green1990,spaute1991,weth1993}.  Because 
$e v_K \approx v_{esc}$, gravitational focusing factors are small and 
growth is slow and orderly \citep{saf1969}. 
The timescale for slow, orderly growth is
\begin{equation}
t_s \approx 30 \left ( \frac{r}{\rm 1000 ~ km} \right ) \left ( \frac{P}{\rm 250 ~ yr} \right ) \left ( \frac{\rm 0.1 ~ g~cm^{-2}}{\Sigma_{\rm 0d}} \right) {\rm Gyr} ~ ,
\label{eq:tslow}
\end{equation}
where $P$ is the orbital period \citep{saf1969,liss1987,gold2004}.

As larger objects form, several processes damp particle random velocities and 
accelerate growth. For objects with $r \sim$ 1--100 m, physical collisions 
reduce particle random velocities \citep{oht1992,kl1998}. For larger
objects with $r \gtrsim$ 0.1 km, the smaller objects damp the orbital 
eccentricity of larger particles through dynamical friction 
\citep{weth1989, kokubo:1995, kl1998}.  Viscous stirring by the large objects 
excites the orbits of the small objects.  For planetesimals with $r \sim$ 
1 m to $r \sim$ 1 km, these processes occur on short timescales, 
$\lesssim 10^6$ yr at 40 AU, and roughly balance when these objects have 
orbital eccentricity $e \sim 10^{-5}$.  
In the case where gas drag is negligible, \citet{gold2004} derive a simple 
relation for the ratio of the eccentricities of the large (`l') and the small 
(`s') objects in terms of their surface densities $\Sigma_{l,s}$ 
\citep[see also][]{kokubo:2002,raf2003a,raf2003b,raf2003c,raf2003d},
\begin{equation}
\frac{e_l}{e_s} \sim \left ( \frac{\Sigma_l}{\Sigma_s} \right )^\gamma ~ ,
\label{eq:dynfric}
\end{equation}
with $\gamma =$ 1/4 to 1/2. Initially, most of the mass is in small 
objects. Thus $\Sigma_l / \Sigma_s \ll$ 1. For $\Sigma_l / \Sigma_s \sim$ 
$10^{-3}$ to $10^{-2}$, $e_l/e_s \approx$ 0.1--0.25.  Because
$e_s v_K \ll$ $v_{l,esc}$ gravitational focusing factors for large 
objects accreting small objects are large.  Runaway growth begins.

Runaway growth relies on positive feedback between accretion and dynamical
friction.  Dynamical friction produces the largest $f_g$ for the largest 
objects, which grow faster and faster relative to the smaller objects and 
contain an ever growing fraction of the total mass. As they grow, these
protoplanets stir the planetesimals. The orbital velocity dispersions of 
small objects gradually approach the escape velocities of the protoplanets. 
With $e_s v_K \sim v_{l,esc}$, collision rates decline as runaway growth 
continues (eqs. (\ref{eq:cross}) and (\ref{eq:dynfric})).
The protoplanets and leftover planetesimals then enter the oligarchic phase, 
where the largest objects -- oligarchs -- grow more slowly than they did as
runaway objects but still faster than the leftover planetesimals.  The 
timescale to reach oligarchic growth is \citep{liss1987,gold2004}
\begin{equation}
t_o \approx 30 \left ( \frac{P}{\rm 250 ~ yr} \right ) \left ( \frac{\rm 0.1 ~ g~cm^{-2}}{\Sigma_{\rm 0d}} \right) {\rm Myr} ~ ,
\label{eq:tfast}
\end{equation}
For the MMSN, $t_o \propto a^{-3}$.  Thus, collisional damping, dynamical 
friction and gravitational focusing enhance the growth rate by 3 orders 
of magnitude compared to orderly growth.

Among the oligarchs, smaller oligarchs grow the fastest.  Each oligarch 
tries to accrete material in an annular `feeding zone' set by balancing 
the gravity of neighboring oligarchs.  If an oligarch accretes all the 
mass in its feeding zone, it reaches the `isolation mass'
\citep{liss1987,kokubo:1998,kokubo:2002,raf2003d,gold2004},
\begin{equation}
m_{iso} \approx 28 \left ( \frac{a}{\rm 40 ~ AU} \right )^3 \left ( \frac{\Sigma_{\rm 0d}}{\rm 0.1 ~ g~cm^{-2}} \right ) M_{\oplus} ~ .
\label{eq:miso}
\end{equation}
Each oligarch stirs up leftover planetesimals along its orbit. Smaller 
oligarchs orbit in regions with smaller $\Sigma_l/\Sigma_s$. Thus, 
smaller oligarchs have larger gravitational focusing factors 
(eqs. (\ref{eq:cross}) and (\ref{eq:dynfric})) and grow faster than 
larger oligarchs \citep{kokubo:1998,gold2004}.

As oligarchs approach $m_{iso}$, they stir up the velocities of the
planetesimals to the disruption velocity. Instead of mergers, collisions 
then yield smaller planetesimals and debris. Continued disruptive 
collisions lead to a collisional cascade, where leftover planetesimals 
are slowly ground to dust \citep{dohn1969,williams:analytical}. Radiation 
pressure from the central star ejects dust grains with $r \lesssim$ 1--10 
$\mu$m; Poynting-Robertson drag pulls larger grains into the central star 
\citep{burns1979,arty1988,take2001}.  Eventually, planetesimals are 
accreted by the oligarchs or ground to dust.  

To evaluate the oligarch mass required for a disruptive collision, we 
consider two planetesimals with equal mass $m_p$. The center-of-mass 
collision energy is 
\begin{equation}
Q_i = v_i^2/8 ~ ,
\label{eq:qimp}
\end{equation}
where the impact velocity $v_i^2 = v^2 + v_{esc}^2$ \citep{weth1993}.  
The energy needed 
to remove half of the combined mass of two colliding planetesimals is
\begin{equation}
Q_D^* = Q_b \left ( \frac{r}{\rm 1 ~ cm} \right)^{\beta_b} + \rho Q_g \left
( \frac{r}{\rm 1 ~ cm} \right)^{\beta_g} ~ ,
\label{eq:qdis}
\end{equation}
where $Q_b r^{\beta_b}$ is the bulk (tensile) component of the binding energy 
and $\rho Q_g r^{\beta_g}$ is the gravity component of the binding energy 
\citep{davis:vesta,housen1990,housen1999,hols1994,benz1999}.  We adopt
$v \approx v_{esc,o}$, where $v_{esc,o} = (G m_o/r_o)^{1/2}$ is the escape 
velocity of an oligarch with mass $m_o$ and radius $r_o$.  We define the
disruption mass $m_d$ by deriving the oligarch mass where $Q_i \approx Q_D^*$.
For icy objects at 30 AU 
\begin{equation}
m_d \sim 3 \times 10^{-6} \left ( \frac{Q_D^*}{10^7 ~ \rm erg~g^{-1}} \right )^{3/2} ~ M_{\oplus}.
\label{eq:mdis}
\end{equation}

\begin{figure}[t] \epsscale{1.05}
\plotone{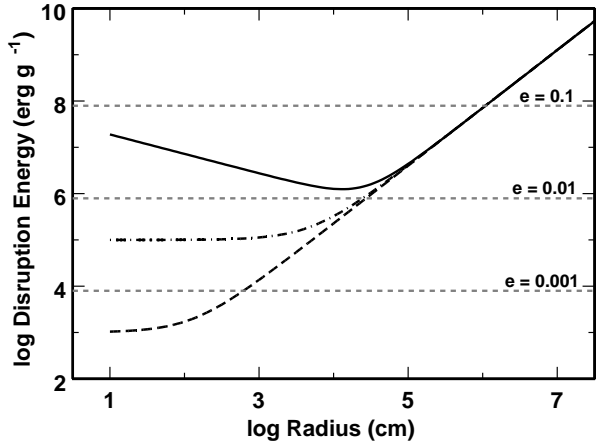}
\vskip -6ex
\caption{ \small 
Disruption energy, $Q_D^*$, for icy objects. The solid curve plots a typical 
result derived from numerical simulations of collisions that include
a detailed equation of state for crystalline ice \citep[$Q_b$
$ = 1.6 \times 10^7$ erg g$^{-1}$, $\beta_b$ = $-$0.42, 
$\rho$ = 1.5 g cm$^{-3}$, $Q_g$ = 1.5 erg cm$^{-3}$, and 
$\beta_g$ = 1.25;][]{benz1999}. The other curves plot results using $Q_b$
consistent with model fits to comet breakups \citep[$\beta_b \approx$ 0;
$Q_b \sim 10^3$ erg g$^{-1}$, dashed curve;
$Q_b \sim 10^5$ erg g$^{-1}$, dot-dashed curve;][]{asp1996}.
The dashed horizontal lines indicate the center of mass collision energy 
(eq. (\ref{eq:qimp})) for equal mass objects with $e$ = 0.001, 0.01, and 
0.1.  Collisions between objects with $Q_i \ll Q_D^*$ yield merged remnants; 
collisions between objects with $Q_i \gg Q_D^*$ produce debris.
}  
\label{fig:qdis}
\end{figure}

Figure \ref{fig:qdis} illustrates the variation of $Q_D^*$ with radius
for several variants of eq. (\ref{eq:qdis}). For icy objects, detailed
numerical collision simulations yield $Q_b \lesssim 10^7$ erg g$^{-1}$, 
$-0.5 \lesssim \beta_b \lesssim$ 0, $\rho \approx$ 1--2 g cm$^{-3}$, 
$Q_g \approx$ 1--2 erg cm$^{-3}$, and $\beta_g$ $\approx$ 1--2 
\citep[solid line in Fig. 1,][see also {\em Chapter by Leinhardt et al.}]
{benz1999}). Models for the breakup of comet Shoemaker-Levy 9 suggest 
a smaller component of the bulk strength, $Q_b \sim 10^3$ erg g$^{-1}$ 
\citep[e.g.,][]{asp1996}, which yields smaller disruption energies for
smaller objects (Fig. 1, dashed and dot-dashed curves). Because nearly
all models for collisional disruption yield similar results for objects
with $r \gtrsim$ 1 km \citep[e.g.,][]{kb2004c}, the disruption mass is
fairly independent of theoretical uncertainties once planetesimals 
become large. For typical $Q_D^* \sim 10^7$--$10^8$ erg g$^{-1}$ for
1--10 km objects (Fig. \ref{fig:qdis}), leftover planetesimals start 
to disrupt when oligarchs have radii, $r_o \sim$ 200--500 km.

Once disruption commences, the final mass of an oligarch depends on the 
timescale for the collisional cascade \citep{kb2004a,kb2004b,kb2004c,lein2005}.
If disruptive collisions produce dust grains much faster than oligarchs 
accrete leftover planetesimals, oligarchs with mass $m_o$ cannot grow 
much larger than 
the disruption radius (maximum oligarch mass $m_{o,max} \approx m_d$). 
However, if oligarchs accrete grains and leftover planetesimals effectively, 
oligarchs reach the isolation mass before collisions and radiation pressure
remove material from the disk \citep[eq. (\ref{eq:miso});][]{gold2004}. The 
relative rates of accretion and disruption depend on the balance between 
collisional damping 
and gas drag -- which slow the collisional cascade -- and viscous stirring 
and dynamical friction -- which speed up the collisional cascade. Because 
deriving accurate rates for these processes requires numerical simulations 
of planetesimal accretion, we now consider simulations of planet formation 
in the Kuiper belt.

\section{\textbf{COAGULATION SIMULATIONS}}

\begin{figure*}[th] \epsscale{2.25}
\plottwo{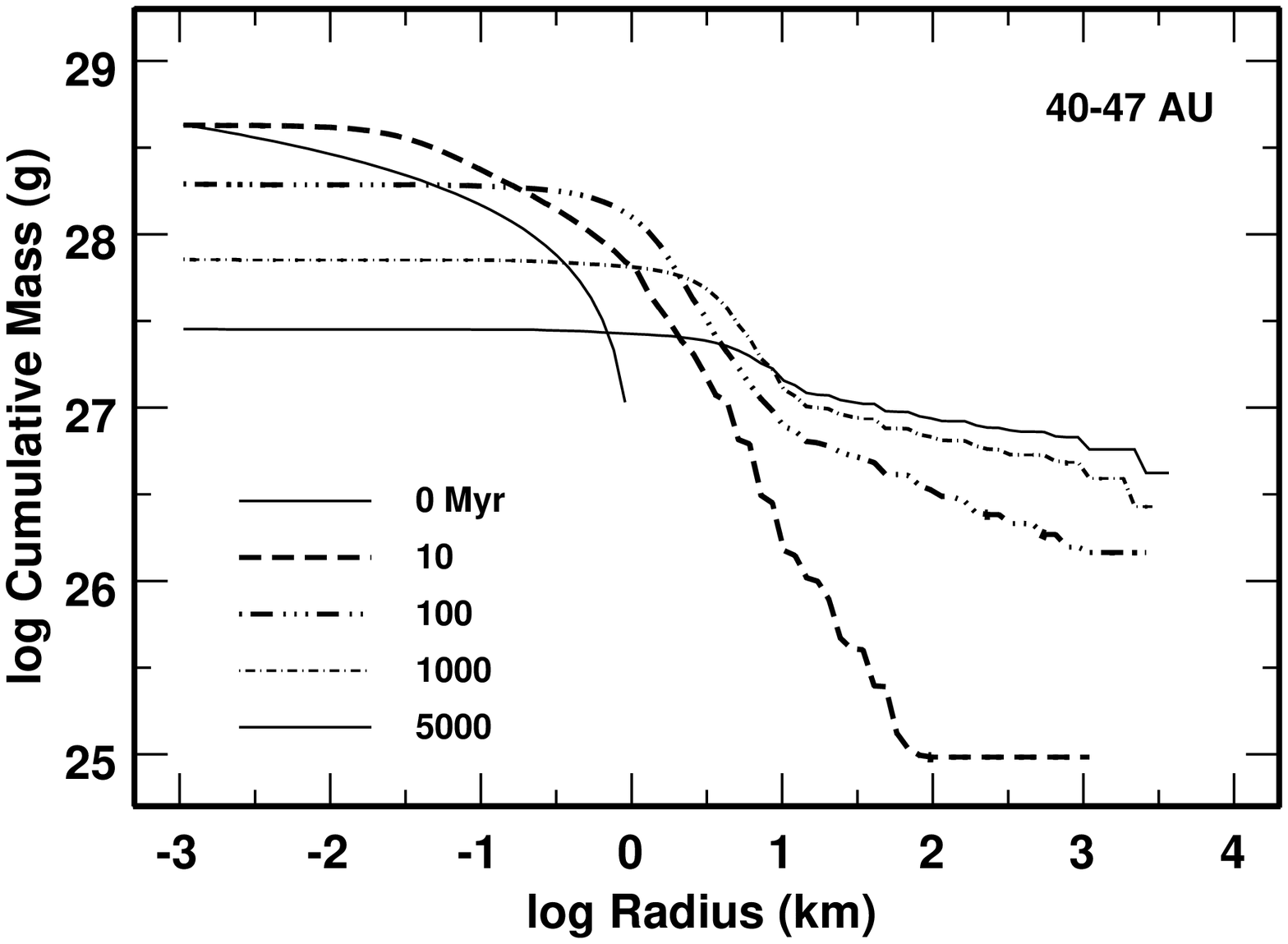}{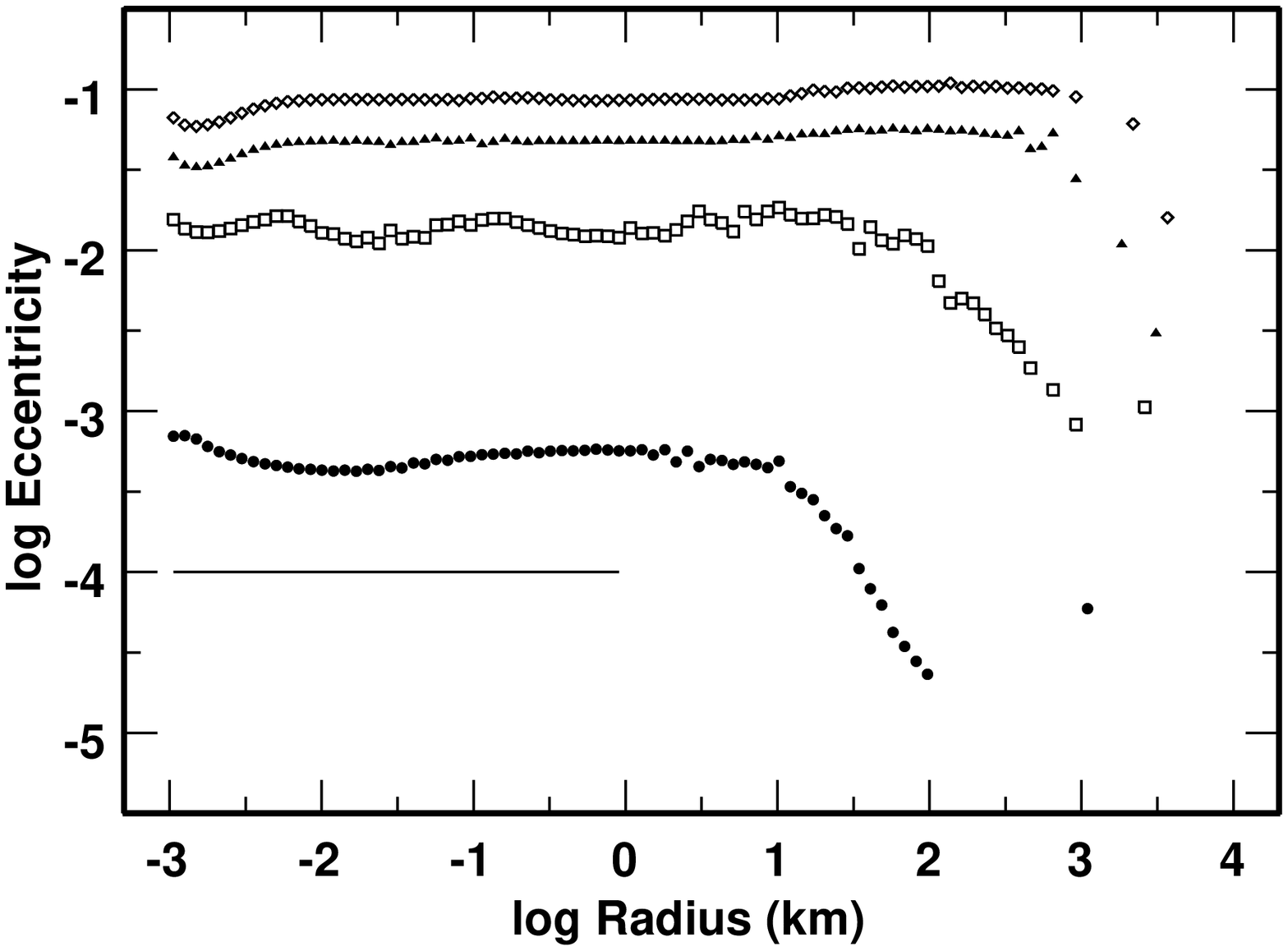}
\vskip -6ex
\caption{ \small 
Evolution of a multiannulus coagulation model with
$\Sigma = 0.12 (a_i/{\rm 40~AU})^{-3/2}$ g cm$^{-2}$.
{\it Left}: cumulative mass distribution at times 
indicated in the legend.
{\it Right}: eccentricity distributions at
$t$ = 0 (light solid line),
$t$ = 10 Myr (filled circles),
$t$ = 100 Myr (open boxes),
$t$ = 1 Gyr (filled triangles), and
$t$ = 5 Gyr (open diamonds).
As large objects grow in the disk, they stir up
the leftover planetesimals to $e \sim$ 0.1. 
Disruptive collisions then deplete the population
of 0.1--10 km planetesimals, which limits the growth
of the largest objects.
}  
\label{fig:massdist}
\end{figure*}

\subsection{Background}

\citet{saf1969} invented the current approach to planetesimal accretion 
calculations.  In his particle-in-a-box method, planetesimals are a 
statistical ensemble of masses with distributions of orbital eccentricities 
and inclinations \citep{green1978,weth1989,weth1993,
spaute1991}.
This statistical approximation is essential: $N$-body codes cannot follow 
the $n \sim 10^9$--$10^{12}$ 1 km planetesimals required to build 
Pluto-mass or Earth-mass planets. For large numbers of objects on fairly 
circular orbits (e.g., $n \gtrsim 10^4$, $r \lesssim$ 1000 km, and 
$e \lesssim$ 0.1), the method is also accurate.  With a suitable prescription 
for collision outcomes, solutions to the coagulation equation in the kinetic 
theory yield the evolution of $n(m)$ with arbitrarily small errors
\citep[e.g.,][]{weth1990,lee2000,mal2001}.  

In addition to modeling planet growth, the statistical approach provides 
a method for deriving the evolution of orbital elements for large ensembles
of planetesimals.  If we (i) assume the distributions of $e$ and $i$ for 
planetesimals follow a Rayleigh distribution and (ii) treat their motions 
as perturbations of a circular orbit, we can use the Fokker-Planck equation 
to solve for small changes in the orbits due to gas drag, gravitational 
interactions, and physical collisions \citep{horn1985,weth1993,ohtsuki:2002}. 
Although the Fokker-Planck equation cannot derive accurate orbital 
parameters for planetesimals and oligarchs near massive planets, it yields 
accurate solutions for the ensemble average $e$ and $i$ when orbital 
resonances and other dynamical interactions are not important 
\citep[e.g.,][]{weth1993,weid1997,ohtsuki:2002}.

Several groups have implemented Safronov's method for calculations
relevant to the outer solar system \citep{green1984,stern1995,stern2005,
stcol1997a,stcol1997b,davis:kbo,kl1998,kl1999a,kl1999b,davis1999,
kb2004a,kb2004c,kb2005}.
These calculations adopt a disk geometry and divide the disk into
$N$ concentric annuli with radial width $\Delta a_i$ at distances
$a_i$ from the central star. Each annulus is seeded with a set
of planetesimals with masses $m_{ij}$, eccentricities $e_{ij}$, and 
inclinations $i_{ij}$, where the index $i$ refers to one of $N$ annuli 
and the index $j$ refers to one of $M$ mass batches within an annulus. 
The mass batches have mass spacing $\delta \equiv m_{j+1}/m_j$.  In most 
calculations, $\delta \approx$ 2; $\delta \le$ 1.4 is optimal 
\citep{oht1990,weth1993,kl1998}.

Once the geometry is set, the calculations solve a set of coupled
difference equations to derive the number of objects $n_{ij}$, and 
the orbital parameters, $e_{ij}$ and $i_{ij}$, as functions of time. 
Most studies allow fragmentation and velocity evolution through 
gas drag, collisional damping, dynamical friction and viscous stirring.  
Because $Q_D^*$ sets the 
maximum size $m_{c,max}$ of objects that participate in the collisional 
cascade, the size distribution for objects with $m < m_{c,max}$ depends 
on the fragmentation parameters \citep[eq. (\ref{eq:qdis});][]{davis:kbo,
kb2004c,pan:2005}.  The size and velocity distributions of the merger 
population with $m > m_{c,max}$ are established during runaway growth 
and the early stages of oligarchic growth. Accurate treatment of velocity 
evolution is important for following runaway growth and thus deriving 
good estimates for the growth times and the size and velocity distributions 
of oligarchs.

When a few oligarchs contain most of the mass, collision rates depend on the 
orbital dynamics of individual objects instead of ensemble averages. Safronov's 
statistical approach then fails \citep[e.g.,][]{weth1993,weiden1997}. 
Although N-body methods can treat the evolution of the oligarchs, they cannot 
follow the evolution of leftover planetesimals, where the statistical approach
remains valid \citep[e.g.,][]{weiden1997}. \citet{spaute1991} solve this 
problem by adding a Monte Carlo treatment of binary interactions between large 
objects to their multiannulus coagulation code. \citet{bk2006} describe a hybrid 
code, which merges a direct N-body code with a multiannulus coagulation code. 
Both codes have been applied to terrestrial planet formation, but not 
to the Kuiper belt. 

Current calculations cannot follow collisional growth accurately in an 
entire planetary system.  Although the six order of magnitude change in 
formation timescales from $\sim$ 0.4 AU to 40 AU is a factor in this
statement, most modern supercomputers cannot finish calculations
involving the entire disk with the required spatial resolution on a 
reasonable timescale.  For the Kuiper belt, it is possible to 
perform a suite of calculations in a disk extending from 30--150 AU 
following 1 m and larger planetesimals. These calculations yield 
robust results for the mass distribution as a function of space and 
time and provide interesting comparisons with observations. Although
current calculations do not include complete dynamical interactions
with the giant planets or passing stars (\citep[see, for example,][]{charnoz:2007}, 
sample calculations clearly show the importance of external perturbations 
in treating the 
collisional cascade. We begin with a discussion of self-stirring 
calculations without interactions with the giant planets or passing
stars and then describe results with external perturbers.

\subsection{Self-Stirring}

To illustrate {\it in situ} KBO formation at 40--150 AU, we consider 
a multiannulus calculation with an initial ensemble of 1 m to 1 km 
planetesimals in a disk with $\Sigma_{0d}$ = 0.12 g cm$^{-2}$.
The planetesimals have initial radii of 1 m to 1 km (with equal mass 
per logarithmic bin), $e_0 = 10^{-4}$, $i_0 = e_0/2$, mass density\footnote{
Our choice of mass density is a compromise between pure ice ($\rho$ = 1
g cm$^{-3}$) and the measured density of Pluto \citep[$\rho \approx$ 2 g cm$^{-3}$][]
{null1993}. The calculations are insensitive to factor of two variations in
the mass density of planetesimals.}
$\rho$ = 1.5 g cm$^{-3}$ and fragmentation parameters 
$Q_b = 10^3$ erg g$^{-1}$, $Q_g$ = 1.5 erg cm$^{-3}$, $\beta_b$ = 0, and 
$\beta_g$ = 1.25 \citep[dashed curve in Fig. \ref{fig:qdis};][]{kb2004c,kb2005}.
The gas density also follows a MMSN, with $\Sigma_g / \Sigma_d$ = 100
and a vertical scale height $h = 0.1 ~ r^{9/8}$ \citep{kh1987}.  The gas 
density is $\Sigma_g \propto e^{-t/t_g}$, with $t_g$ = 10 Myr.

This calculation uses an updated version of the \citet{bk2006} code that 
includes a Richardson extrapolation procedure in the coagulation algorithm. 
As in the Eulerian \citep{kl1998} and fourth order Runge-Kutta 
\citep{kb2002a,kb2002b} methods employed previously, this code provides
robust numerical solutions to kernels with analytic solutions 
\citep[e.g.,][]{oht1990,weth1990} without producing the wavy size 
distributions described in other simulations with a low mass cutoff
\citep[e.g.][]{campo:waves}. Once the evolution of large ($r >$ 1 m)
objects is complete, a separate code tracks the evolution of lower
mass objects and derives the dust emission as a function of time.

Figure \ref{fig:massdist} shows the evolution of the mass and 
eccentricity distributions at 40--47 AU for this calculation. During 
the first few Myr, the largest objects grow slowly. Dynamical friction 
damps the orbits of the largest objects; collisional damping and gas 
drag circularize the orbits of the smallest objects. This evolution 
erases many of the initial conditions and enhances gravitational 
focusing by factors of 10--1000. Runaway growth begins. A few (and 
sometimes only one) oligarchs then grow from $r \sim$ 10 km to $r \sim$ 
1000 km in $\sim$ 30 Myr at 40 AU and in $\sim$ 1 Gyr at 150 AU (see 
eq. (\ref{eq:tfast})).
Throughout runaway growth, dynamical friction and viscous stirring raise 
the random velocities of the leftover planetesimals to $e \approx$ 0.01--0.1 
and $i \approx$ 2$^{\rm o}$--4$^{\rm o}$
($v \sim$ 50--500 m s$^{-1}$ at 40--47 AU; Figure \ref{fig:massdist}; 
right panel).  Stirring reduces gravitational focusing factors and 
ends runaway growth. The large oligarchs then grow slowly through 
accretion of leftover planetesimals.

As oligarchs grow, collisions among planetesimals initiate the 
collisional cascade. Disruptive collisions dramatically reduce the 
population of 1--10 km objects, which slows the growth of oligarchs 
and produces a significant debris tail in the size distribution. In 
these calculations, disruptive collisions remove material from the 
disk faster than oligarchs can accrete the debris. Thus, growth 
stalls and produces $\sim$ 10--100 objects with maximum sizes 
$r_{max} \sim$ 1000--3000 km at 40--50 AU 
\citep{stcol1997a,stcol1997b,kb2004c,kb2005,stern2005}.

Stochastic events lead to large dispersions in the growth time for
oligarchs, $t_o$ (eq. (\ref{eq:tfast})). In ensembles of 25--50 simulations 
with identical starting conditions, an occasional oligarch will grow up to 
a factor of two faster than its neighbors. This result occurs in 
simulations with $\delta$ = 1.4, 1.7, and 2.0, and thus seems independent 
of mass resolution. These events occur in $\sim$ 25\% of the simulations 
and lead to factor of $\sim$ 2 variations in $t_o$ (eq. (\ref{eq:tfast})).

\begin{figure}[t] \epsscale{1.25}
\plotone{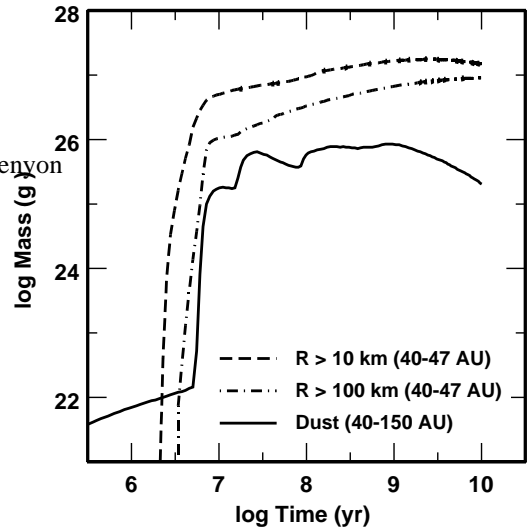}
\vskip -7ex
\caption{ \small
Time evolution of the mass in KBOs and dust grains.
Solid line: dust mass ($r \lesssim$ 1 mm) at 40--150 AU.
Dashed (dot-dashed) lines: total mass in small (large) KBOs 
at 40--47 AU.
}
\label{fig:dustevol}
\end{figure}

In addition to modest-sized icy planets, oligarchic growth generates 
copious amounts of dust (Figure \ref{fig:dustevol}). When runaway growth 
begins, collisions produce small amounts of dust from `cratering' 
\citep[see, for example][]{green1978,weth1993,stcol1997a,stcol1997b,kl1999a}.
Stirring by growing oligarchs leads to `catastrophic' collisions, where 
colliding planetesimals lose more than 50\% of their initial mass. These 
disruptive collisions produce a spike in the dust production rate that 
coincides with the formation of oligarchs with $r \gtrsim$ 200--300 km 
(eq. (\ref{eq:mdis})).  
As the wave of runaway growth propagates outward, stirring produces 
disruptive collisions at ever larger heliocentric distances. The dust 
mass grows in time and peaks at $\sim$ 1 Gyr, when oligarchs reach 
their maximum mass at 150 AU. As the mass in leftover planetesimals 
declines, Poynting-Robertson drag removes dust faster than disruptive 
collisions produce it. The dust mass then declines with time.

\subsection{External Perturbation}

Despite the efficiency of self-stirring models in removing leftover
planetesimals from the disk, other mechanisms must reduce the derived 
mass in KBOs to current observational limits. In 
self-stirring calculations at 40--50 AU, the typical mass in KBOs with 
$r \sim$ 30---1000 km at 4--5 Gyr is a factor of 5--10 larger than 
currently observed \citep[][{\em Chapter by Petit et al.}]{luu2002}. 
Unless Earth-mass or larger objects form in the Kuiper belt 
\citep{chiang2007,lev2007}, external perturbations must
excite KBO orbits and enhance the collisional cascade.

Two plausible sources of external perturbation can reduce the 
predicted KBO mass to the desired limits. Once Neptune achieves
its current mass and orbit, it stirs up the orbits of KBOs at 35--50 AU 
\citep{levi1990,holman:1993,duncan1995,kuchner2002, morby2004b}.  In 
$\sim$ 100 Myr or less, Neptune removes nearly all KBOs with $a \lesssim$ 
37--38 AU.  Beyond $a \sim$ 38 AU, some KBOs are trapped in 
orbital resonance with Neptune \citep{malhotra:1995,malhotra:1996};
others are ejected into the scattered disk \citep{duncan:1997}.
In addition to these processes, Neptune stirring increases the 
effectiveness of the collisional 
cascade \citep{kb2004c}, which removes additional mass from the 
population of 0.1--10 km KBOs and prevents growth of larger KBOs. 

Passing stars can also excite KBO orbits and enhance the collisional 
cascade.  Although Neptune dynamically ejects scattered disk objects 
with perihelia $q \lesssim $ 36--37 AU \citep{morby2004b},
objects with $q \gtrsim$ 45--50 AU, such as Sedna and Eris, require 
another scattering source. Without evidence for massive planets at 
$a \gtrsim$ 50 AU \citep{morbidelli:2002}, a passing star is the most 
likely source of the large $q$ for these KBOs \citep{morby2004a,kb2004d}.

\citet[][see also {\em Chapter by Duncan et al.}]{adams2001} examined the probability of encounters between the 
young Sun and other stars. Most stars form in dense clusters with 
estimated lifetimes of $\sim$ 100 Myr. To account for the abundance 
anomalies of radionuclides in solar system meteorites (produced by 
supernovae in the cluster) and for the stability of Neptune's orbit 
at 30 AU, the most likely solar birth cluster has $\sim$ 2000--4000 
members, a crossing time of $\sim$ 1 Myr, and a relaxation time of 
$\sim$ 10 Myr.  The probability of a close encounter with a distance
of closest approach $a_{close}$ is then $\sim$ 60\% 
($a_{close}$/160 AU)$^2$ \citep{kb2004d}.

Because the dynamical interactions between KBOs in a coagulation 
calculation and large objects like Neptune or a passing star are 
complex, here we consider simple calculations of each process.  To 
illustrate the evolution of KBOs after a stellar flyby, we consider
a very close pass with $a_{close}$ = 160 AU \citep{kb2004d}.  This 
co-rotating flyby produces objects with orbital parameters similar 
to those of Sedna and Eris.  For objects in the coagulation 
calculation, the flyby produces an $e$ distribution 
\begin{equation}
e_{KBO} = \left\{ \begin{array}{l l l}
        0.025 (a / {\rm 30~AU})^4 & \hspace{4mm} & a < a_0 \\
\\
        0.5 & \hspace{4mm} & a > a_0 \\
         \end{array}
         \right .
\label{eq:eccflyby}
\end{equation}
with $a_0 \approx$ 65 AU \citep[see][]{ida2000,kb2004d,koba2005}.
This $e$ distribution produces a dramatic increase in the debris 
production rate throughout the disk, which freezes the mass distribution 
of the largest objects\footnote{ The $i$ distribution following a flyby
depends on the relative orientations of two planes, the orbital plane
of KBOs and the plane of the trajectory of the passing star. Here, we
assume the flyby produces no change in $i$, which simplifies the
discussion without changing any of the results significantly. 
}.
Thus, to produce an ensemble of KBOs with $r \gtrsim$ 300 km at 
40--50 AU, the flyby must occur when the Sun has an age 
$t_{\odot} \gtrsim$ 10--20 Myr (Figure \ref{fig:massdist}).
For $t_{\odot} \gtrsim$ 100 Myr, the flyby is very unlikely. As
a compromise between these two estimates, we consider a flyby at
$t_{\odot} \sim$ 50 Myr.

\begin{figure}[ht] \epsscale{1.0}
\plotone{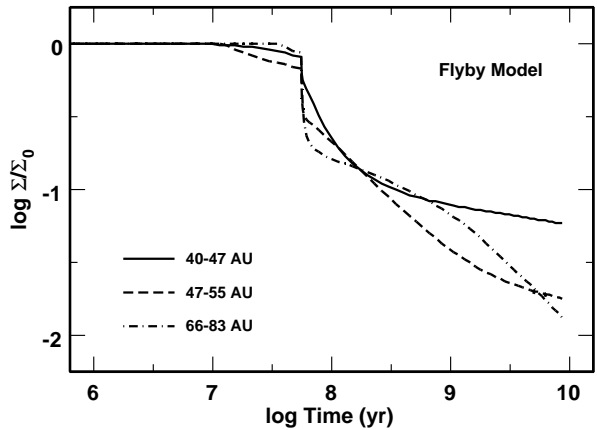}
\vskip -6ex
\caption{ \small 
Evolution of $\Sigma$ after a stellar flyby. After 50 Myr of growth,
the close pass excites KBOs to large $e$ (eq. (\ref{eq:eccflyby}))
and enhances the collisional cascade. 
}  
\label{fig:flybysig}
\end{figure}

Figure \ref{fig:flybysig} shows the evolution of the KBO surface 
density in three annuli as a function of time. At early times 
($t \lesssim$ 50 Myr), KBOs grow in the standard way.  After the 
flyby, the disk suffers a dramatic loss of material. At 40--47 AU,
the disk loses $\sim$ 90\% (93\%) of its initial mass in $\sim$ 
1 Gyr (4.5 Gyr). At $\sim$ 50--80 AU, the collisional cascade removes
$\sim$ 90\% (97\%) of the initial mass in $\sim$ 500 Myr (4.5 Gyr).
Beyond $\sim$ 80 AU, KBOs contain less than 1\% of the initial mass.
Compared to self-stirring models, flybys that produce Sedna-like orbits 
are a factor of 2--3 more efficient at removing KBOs from the solar 
system.

To investigate the impact of Neptune on the collisional cascade, we 
parameterize the growth of Neptune at 30 AU as a simple function of 
time \citep{kb2004c}
\begin{equation}
M_{Nep} \approx \left\{ \begin{array}{l l l}
        6 \times 10^{27} ~ e^{(t-t_N)/t_1} ~ {\rm g} & \hspace{1mm} & t < t_N \\
\\
        6 \times 10^{27} ~ {\rm g} ~ + ~ C(t-t_1) & \hspace{1mm} & t_N < t
 < t_2 \\
\\
        1.0335 \times 10^{29} ~ {\rm g} & \hspace{3mm} & t > t_2 \\
         \end{array}
         \right .
\label{eq:Nepmass}
\end{equation}
where $C_{Nep}$ = $1.947 \times 10^{21}$ g yr$^{-1}$, $t_N$ = 50 Myr, 
$t_1$ = 3 Myr, and $t_2$ = 100 Myr.  These choices enable a model 
Neptune to reach a mass of 1 $M_{\oplus}$ in 50 Myr, when the largest
KBOs form at 40--50 AU, and reach its current mass in 100 Myr\footnote{
This prescription is not intended as an accurate portrayal of Neptune 
formation, but it provides a simple way to investigate how Neptune might
stir the Kuiper belt once massive KBOs form.}.  As Neptune 
approaches its final mass, its gravity stirs up KBOs at 40--60 AU
and increases their orbital eccentricities to $e \sim$ 0.1--0.2 on short 
timescales.  In the coagulation model, distant planets produce negligible
changes in $i$, so self-stirring sets $i$ in these calculations 
\citep{weid1989}.  This evolution enhances debris production by a factor of 
3--4, which effectively freezes the mass distribution of 100--1000 km 
objects at 40--50 AU.  By spreading the leftover planetesimals and the
debris over a larger volume, Neptune stirring limits the growth of the
oligarchs and thus reduces the total mass in KBOs.

\begin{figure}[t] \epsscale{1.0}
\plotone{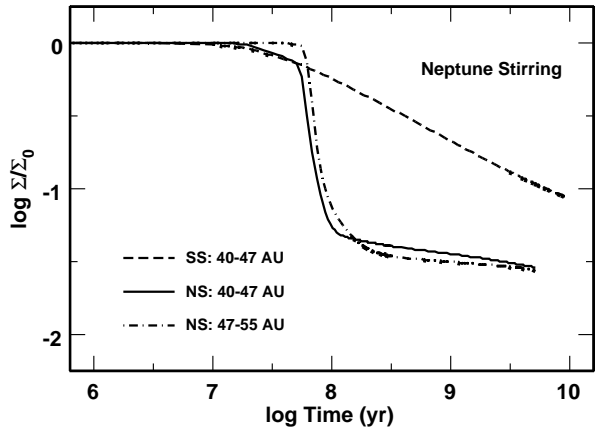}
\vskip -6ex
\caption{ \small
Evolution of $\Sigma$(KBO) in models with Neptune stirring.
Compared to self-stirring models (SS; dashed curve), stirring by
Neptune rapidly removes KBOs at 40--47 AU (NS; solid cruve) and 
at 47--55 AU (NS; dot-dashed curve). 
}
\label{fig:Nepsig}
\end{figure}

Figure \ref{fig:Nepsig} shows the evolution of the surface density
in small and large KBOs in two annuli as a function of time.  At 
40--55 AU, Neptune rapidly stirs up KBOs to $e \sim$ 0.1 when it 
reaches its current mass at $\sim$ 100 Myr. Large collision velocities 
produce more debris, which is rapidly ground to dust and removed from 
the system by radiation pressure at early times and by Poynting-Robertson
drag at later times. Compared to self-stirring models, the change in 
$\Sigma$ is dramatic, with only $\sim$ 3\% of the initial disk mass 
remaining at $\sim$ 4.5 Gyr.

\begin{figure}[th] \epsscale{1.0}
\plotone{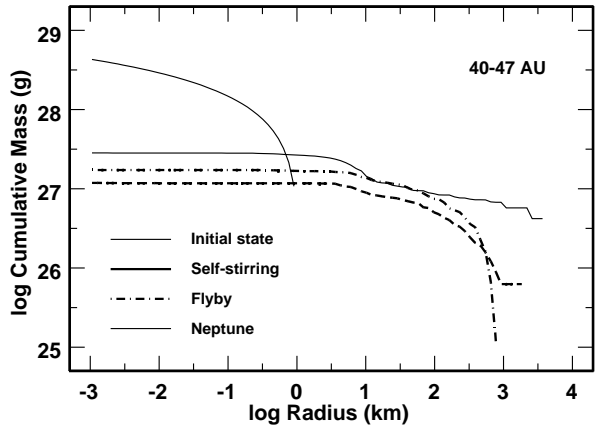}
\vskip -6ex
\caption{ \small 
Mass distributions for evolution with self-stirring (heavy solid line), 
stirring from a passing star (dot-dashed line), and stirring from Neptune 
at 30 AU (dashed line).  After 4.5 Gyr, the mass in KBOs with $r \gtrsim$ 
50 km is $\sim$ 5\% (self-stirring), $\sim$ 3.5\% (flyby), and $\sim$ 2\% 
(Neptune stirring) of the initial mass. The number of objects with $r \gtrsim$
1000 km is $\sim$ 100 (self-stirring), $\sim$ 1 (flyby), and $\sim$ 10
(Neptune stirring). The largest object has $r_{max} \sim$ 3000 km 
(self-stirring), $r_{max} \sim$ 500--1000 km (flyby), and 
$r_{max} \sim$ 1000--2000 km (Neptune stirring). 
}  
\label{fig:allmass}
\end{figure}

From these initial calculations, it is clear that external 
perturbations dramatically reduce the mass of KBOs in the disk
\citep[see also][]{charnoz:2007}.
Figure \ref{fig:allmass} compares the mass distributions at
40--47 AU and at 4.5 Gyr for the self-stirring model in Figure 
\ref{fig:massdist} (solid line) with results for the flyby
(dot-dashed line) and Neptune stirring (dashed line).  Compared 
to the self-stirring model, the close flyby reduces the mass in
KBOs by $\sim$ 50\%. Neptune stirring reduces the KBO mass by
almost a factor of 3 relative to the self-stirring model. For 
KBOs with $r \gtrsim$ 30--50 km, the predicted mass in KBOs with 
Neptune stirring is within a factor of 2--3 of the current mass 
in KBOs. 

These simple calculations for the stellar flyby and Neptune stirring
do not include dynamical depletion. In the stellar flyby picture, 
the encounter removes nearly all KBOs beyond a truncation radius,
$a_T \sim$ 48 ($a_{close}$ / 160 AU) AU \citep{kb2004d}. Thus, a close 
pass with $a_{close}$ $\sim$ 160 AU can produce the observed outer edge 
of the Kuiper belt at 48 AU.  Although many objects with initial $a > a_T$ 
are ejected from the Solar System, some are placed on very elliptical, 
Sedna-like orbits\footnote{\citet{levi2004} consider the impact of the 
flyby on the scattered disk and Oort cloud. After analyzing a suite of
numerical simulations, they conclude that the flyby must occur before 
Neptune reaches its current orbit and begins the dynamical processes that 
populate the Oort cloud and the scattered disk. If Neptune forms {\it in situ} 
in 1--10 Myr, then the flyby cannot occur after massive KBOs form. If
Neptune migrates to 30 AU after massive KBOs form, then a flyby can
truncate the Kuiper belt without much impact on the Oort cloud or the
scattered disk.}.
In the Neptune stirring model, dynamical interactions 
will eject some KBOs at 40--47 AU.  If the dynamical interactions that 
produce the scattered disk reduce the mass in KBOs by a factor of 2 at 
40--47 AU \citep[e.g.,][]{duncan1995,kuchner2002}, the Neptune stirring 
model yields a KBO mass in reasonably good agreement with observed limits
\citep[for a different opinion, see][]{charnoz:2007}.

\subsection{Nice Model}

Although {\it in situ} KBO models can explain the current amount of 
mass in large KBOs, these calculations do not address the orbits of 
the dynamical classes of KBOs. To explain the orbital architecture 
of the giant planets, the `Nice group' centered at Nice Observatory
developed an inspired, sophisticated picture of the dynamical evolution 
of the giant planets and a remnant planetesimal disk
\citep[][and references therein]{tsig2005,morby2005,gomes2005}. 
The system begins in an approximate equilibrium, with the 
giant planets in a compact configuration (Jupiter at 5.45 AU, 
Saturn at $\sim$ 8.2 AU, Neptune at $\sim$ 11.5 AU, and Uranus 
at $\sim$ 14.2 AU) and a massive planetesimal disk at 15--30 AU.
Dynamical interactions between the giant planets and the planetesimals 
lead to an instability when Saturn crosses the 2:1 orbital resonance with 
Jupiter, which results in a dramatic orbital migration of the gas giants 
and the dynamical ejection of planetesimals into the Kuiper belt, 
scattered disk, and the Oort cloud. Comparisons between the end state of 
this evolution and the orbits of KBOs in the `hot population' and the 
scattered disk are encouraging ({\em Chapter by Morbidelli et al.}).

Current theory cannot completely address the likelihood of the initial 
state in the Nice model. 
\citet{thom1999,thom2002} demonstrate that $n$-body simulations can produce 
a compact configuration of gas giants, but did not consider how 
fragmentation or interactions with low mass planetesimals affect 
the end state. \citet{obrien2005} show that a disk of planetesimals
has negligible collisional grinding over 600 Myr if most of the mass
is in large planetesimals with $r \gtrsim$ 100 km. However, they did
not address whether this state is realizable starting from an ensemble
of 1 km and smaller planetesimals.  In terrestrial planet simulations
starting with 1--10 km planetesimals, the collisional cascade removes 
$\sim$ 25\% of the initial rocky material in the disk 
\citep{weth1993,kb2004b}. Interactions between oligarchs and remnant 
planetesimals are also important for setting the final mass and 
dynamical state of the terrestrial planets \citep{bk2006,kb2006}.
Because complete hybrid calculations of the giant planet region are 
currently computationally prohibitive, it is not possible to
make a reliable assessment of these issues for the formation of
gas giant planets.

\begin{figure}[t] \epsscale{1.0}
\plotone{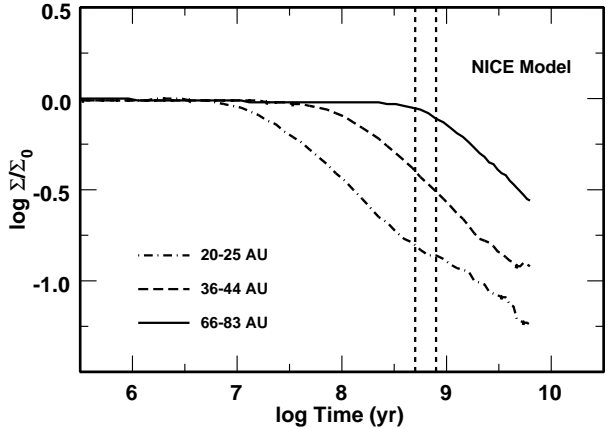}
\vskip -6ex
\caption{ \small
Evolution of $\Sigma$ in a self-stirring model at 20--100 AU.  At 20--25 AU, 
it takes $\sim$ 5--10 Myr to form 1000 km objects. After $\sim$ 0.5--1 Gyr, 
there are $\sim$ 100 objects with $r \sim$ 1000--2000 km and $\sim$ $10^5$ 
objects with $r \sim$ 100--200 km at 20--30 AU.  As these objects grow, the 
collisional cascade removes $\sim$ 90\% of the mass in remnant planetesimals. 
The twin vertical dashed lines bracket the time of the Late Heavy Bombardment 
at $\sim$ 300--600 Myr.
}
\label{fig:nice}
\end{figure}

Here, we consider the evolution of the planetesimal disk outside
the compact configuration of giant planets, where standard 
coagulation calculations can follow the evolution of many initial
states for 1--5 Gyr in a reasonable amount of computer time.  
Figure \ref{fig:nice} shows the time evolution for the surface 
density of planetesimals in three annuli from one typical 
calculation at 20--25 AU (dot-dashed curve; $M_i$ = 6 $M_{\oplus}$), 
36--44 AU (dashed curve; $M_i$ = 9 $M_{\oplus}$), and 66--83 AU 
(solid curve; $M_i$ = 12 $M_{\oplus}$). Starting from the standard 
surface density profile (eq. \ref{eq:mmsn}), planetesimals at 
20--25 AU grow to 1000 km sizes in a few Myr. Once the collisional
cascade begins, the surface density slowly declines to $\sim$ 10\%
to 20\% of its initial value at the time of the Late Heavy Bombardment,
when the Nice model predicts that Saturn crosses the 2:1 orbital
resonance with Jupiter.

These results provide a strong motivation to couple coagulation calculations
with the dynamical simulations of the Nice group \citep[see also][]
{charnoz:2007}. In the Nice model, dynamical interactions with a 
massive planetesimal disk are the `fuel' for the dramatic migration of 
the giant planets and the dynamical ejection of material into the 
Kuiper belt and the scattered disk. If the mass in the planetesimal 
disk declines by $\sim$ 80\% as the orbits of the giant planets evolve, 
the giant planets cannot migrate as dramatically as in the 
{\em Gomes et al.} (2005) calculations. Increasing the initial mass in 
the disk by a factor of 3--10 may allow coagulation and the collisional 
cascade to produce a debris disk capable of triggering the scattering 
events of the Nice model. 

\subsection{A Caveat on the Collisional Cascade}

Although many of the basic outcomes of oligarchic growth and the 
collisional cascade are insensitive to the initial conditions and 
fragmentation parameters for the planetesimals, several uncertainties 
in the collisional cascade can modify the final mass in oligarchs and 
the distributions of $r$ and $e$. Because current computers do not 
allow coagulation calculations that include the full range of sizes 
(1 $\mu$m to $10^4$ km), published calculations have two pieces, a 
solution for large objects \citep[e.g.,][]{kb2004a,kb2004b} and a 
separate solution for smaller objects \citep[e.g.,][]{kriv2006}.
Joining these solutions assumes that (i) collision fragments continue
to collide and fragment until particles are removed by radiative 
processes and (ii) mutual (destructive) collisions among the fragments
are more likely than mergers with much larger oligarchs. These
assumptions are reasonable but untested by numerical calculations
\citep{kb2002a}. Thus, it may be possible to halt or to slow the 
collisional cascade before radiation pressure rapidly remove small 
grains with $r \approx$ 1--100 $\mu$m. 

In current coagulation calculations, forming massive oligarchs at 5--15 AU 
in a massive disk requires an inefficient collisional cascade. When the
cascade is efficient, the most massive oligarchs have $m \lesssim$ 
1 $M_{\oplus}$.  Slowing the cascade allows oligarchs to accrete 
planetesimals more efficiently, which results in larger oligarchs that 
contain a larger fraction of the initial mass. If collisional damping 
is efficient, halting the cascade completely at sizes of $\sim$ 1 mm leads 
to rapid {\it in situ} formation of Uranus and Neptune \citep{gold2004}
and early stirring of KBOs at 40 AU.

There are two simple ways to slow the collisional cascade. In simulations
where the cascade continues to small sizes, $r \sim$ 1--10 $\mu$m, the
radial optical depth in small grains is $\tau_s \sim$ 0.1--1 at 30--50 AU 
\citep{kb2004a}.  Lines-of-sight to the central star are 
not purely radial, so this optical depth reduces radiation pressure and 
Poynting-Robertson drag by small factors, $\sim e^{-0.2 \tau_s}$ $\sim$ 
10\%--30\%, and has little impact on the evolution of the cascade. With
$\tau_s \propto$ $a^{-s}$ and $s \sim$ 1--2, however, the optical depth 
may reduce radiation forces significantly at smaller $a$.  Slowing the 
collisional cascade by factors of 2--3 could allow oligarchs to accrete 
leftover planetesimals and smaller objects before the cascade removes them.

Collisional damping and gas drag on small particles may also slow the
collisional cascade.  For particles with large ratios of surface area
to volume, $r \lesssim$ 0.1--10 cm, collisions and the gas effectively 
damp $e$ and $i$ \citep{ada76,gold2004} and roughly 
balance dynamical friction and viscous stirring. Other interactions between
small particles and the gas -- such as photophoresis \citep{wurm2006} --
also damp particles randome velocities and thus might help to slow the 
cascade. Both collisions and interactions
between the gas and the solids are more effective at large volume density, 
so these processes should be more important inside 30 AU than outside 30 AU. 
The relatively short lifetime of the 
gas, $\sim$ 3--10 Myr, also favors more rapid growth inside 30 AU.  
If damping maintains an equilibrium $e \sim 10^{-3}$ at $a \sim$ 20--30 AU, 
oligarchs can grow to the sizes, $r \gtrsim$ 2000 km, required in the Nice
model. Rapid growth at $a \sim$ 5--15 AU might produce oligarchs with the 
isolation mass ($r \sim$ 10--30 $R_{\oplus}$; eq. \ref{eq:miso}) and lead
to the rapid formation of gas giants. 

Testing these mechanisms for slowing the collisional cascade requires
coagulation calculations with accurate treatments of collisional damping,
gas drag, and optical depth for particle radii $r \sim$ 1--10 $\mu$m to 
$r \sim$ 10000 km. Although these calculations require factors of 4--6
more computer time than published calculations, they are possible with
multiannulus coagulation codes on modern parallel computers.

\subsection{Model Predictions}

The main predictions derived from coagulation models are $n(r)$, $n(e)$,
and $n(i)$ as functions of $a$.  The cumulative number distribution 
consists of three power laws \citep{kb2004c,pan:2005}
\begin{equation}
n(r) = \left\{ \begin{array}{l l l}
	n_d r^{-\alpha_d} & \hspace{5mm} & r \le r_1 \\
\\
  	n_i		  & \hspace{5mm} & r_1 \le r < r_0 \\
\\
  	n_m r^{-\alpha_m} & \hspace{5mm} & r \ge r_0 \\
	 \end{array}
	 \right .
\label{eq:sizedist}
\end{equation}
The debris population at small sizes, $r \le r_1$, always has 
$\alpha_d \approx$ 3.5.  The merger population at large sizes,
$r \ge r_0$, has $\alpha_m \approx$ 2.7--4.  Because the collisional 
cascade robs oligarchs of material, calculations with more stirring 
have steeper size distributions.  Thus, self-stirring calculations 
with $Q_b \gtrsim 10^5$ erg g$^{-1}$ ($Q_b \lesssim 10^3$ erg g$^{-1}$) 
typically yield $\alpha_m \approx$ 2.7--3.3 (3.5--4). Models with a stellar 
flyby or stirring by a nearby gas giant also favor large $\alpha_m$.

The transition radii for the power laws depend on the fragmentation
parameters (see Fig. \ref{fig:qdis}; see also \citet{pan:2005}).  
For a typical $e \sim$ 0.01--0.1
in self-stirring models, $r_0 \approx r_1 \approx$ 1 km 
when $Q_b \gtrsim 10^5$ erg g$^{-1}$. When $Q_b \lesssim 10^3$ erg g$^{-1}$, 
$r_1 \approx$ 0.1 km and $r_0 \approx$ 10--20 km.  Thus the calculations 
predict a robust correlation between the transition radii and the power 
law exponents: large $r_0$ and $\alpha_m$ or small $r_0$ and $\alpha_m$.

Because gravitational stirring rates are larger than accretion rates, the 
predicted distributions of $e$ and $i$ at 4--5 Gyr depend solely on the 
total mass in oligarchs \citep[see also][]{gold2004}. Small objects with 
$r \lesssim r_0$ contain a very small fraction of the mass and cannot stir
themselves. Thus $e$ and $i$ are independent of $r$ (Fig. \ref{fig:massdist}).  
The $e$ and $i$ for larger objects depends on the total mass in the largest
objects. In self-stirring models, dynamical friction and viscous stirring
between oligarchs and planetesimals (during runaway growth) and among the
ensemble of oligarchs (during oligarchic growth) set the distribution of
$e$ for large objects with $r \gtrsim r_0$. In self-stirring models, viscous 
stirring among oligarchs dominates dynamical friction between oligarchs and
leftover planetesimals, which leads to a shallow relation between $e$ and $r$,
$e \propto r^{-\gamma}$ with $\gamma \approx$ 3/4. In the flyby and Neptune
stirring models, stirring by the external perturber dominates stirring among 
oligarchs. This stirring yields a very shallow relation between $e$ and $r$ 
with $\gamma \approx$ 0--0.25.

Other results depend little on the initial conditions and the fragmentation 
parameters. In calculations with different initial mass
distributions, an order of magnitude range in $e_0$, and $Q_b = 
10^0$--$10^7$ erg g$^{-1}$, $\beta_b$ = $-$0.5--0, $Q_g =$ 0.5--5 
erg cm$^{-3}$, and $\beta_g \ge$ 1.25, $r_{max}$ and the amount of 
mass removed by the collisional cascade vary by $\lesssim$ 10\% 
relative to the evolution of the models shown in Figures 
\ref{fig:massdist}--\ref{fig:nice}.  Because collisional damping among 
1 m to 1 km objects erases the initial orbital distribution, the 
results do not depend on $e_0$ and $i_0$. Damping and dynamical
friction also quickly erase the initial mass distribution, which yields
growth rates that are insensitive to the initial mass distribution.

The insensitivity of $r_{max}$ and mass removal to the fragmentation 
parameters depends on the rate of collisional disruption relative to the
growth rate of oligarchs.  Because the collisional cascade starts when 
$m_o \sim m_d$ (eq. (\ref{eq:mdis})), calculations with small $Q_b$ 
($Q_b \lesssim 10^3$ erg g$^{-1}$) produce large amounts of debris before 
calculations with large $Q_b$ ($Q_b \gtrsim 10^3$ erg g$^{-1}$). Thus, 
an effective collisional cascade should yield lower mass oligarchs 
and more mass removal when $Q_b$ is small.  However, oligarchs with 
$m_o \sim m_d$ still have fairly large gravitational focusing factors 
and accrete leftover planetesimals more rapidly than the cascade removes 
them. As oligarchic growth continues, gravitational focusing factors fall 
and collision disruptions increase.  All calculations then reach a point
where the collisional cascade removes leftover planetesimals more rapidly 
than oligarchs can accrete them. As long as most planetesimals have
$r \sim$ 1--10 km, the timing of this epoch is more sensitive to 
gravitational focusing and the growth of oligarchs than the collisional
cascade and the fragmentation parameters. Thus, $r_{max}$ and the amount
of mass processed through the collisional cascade are relatively insensitive
to the fragmentation parameters.

\section{Confronting KBO collision models with KBO data}

Current data for KBOs provide two broad tests of coagulation calculations.
In each dynamical class, four measured parameters test the general results 
of coagulation models and provide ways to discriminate among the outcomes 
of self-stirring and perturbed models. These parameters are

\begin{itemize}

\item $r_{max}$, the size of the largest object, 

\item $\alpha_m$, the slope of the size distribution for 
large KBOs with $r \gtrsim$ 10 km,

\item $r_0$, the break radius, which measures the radius where the 
size distribution makes the transition from a merger population
($r \gtrsim r_0$) to a collisional population ($r \lesssim r_0$)
as summarized in eq. (\ref{eq:sizedist}), and

\item $M_l$, the total mass in large KBOs.

\end{itemize}

\noindent
For all KBOs, 
measurements of the dust mass allow tests of the collisional cascade and 
link the Kuiper belt to observations of nearby debris disks. We begin with 
the discussion of large KBOs and then compare the Kuiper belt with other 
debris disks.

Table 1 summarizes the mass and size distribution parameters derived from
recent surveys.  To construct this table, we used online data from the
Minor Planet Center (http:$\rm //cfa-www.harvard.edu/iau/lists/MPLists.html$)
for $r_{max}$ \citep[see also][]{lev2001} and the results of several detailed 
analyses for $\alpha_m$, $r_{max}$, and $r_0$ \citep[e.g.,][{\em Chapter} 
by {\em Petit et al.}]{bernstein:tnodist,elliot2005,petit:2006}. Because 
comprehensive KBO surveys are challenging, the entries in the Table are
incomplete and sometimes uncertain. Nevertheless, these results provide
some constraints on the calculations.

Current data provide clear evidence for physical differences among the
dynamical classes. For classical KBOs with $a$ = 42--48 AU and $q >$ 37 AU, 
the cold population ($i \lesssim$ 4$^{\rm o}$) has a steep size distribution 
with $\alpha_m \approx$ 3.5--4 and $r_{max} \sim$ 300--500 km.  In contrast, 
the hot population ($i \gtrsim$ 
10$^{\rm o}$) has a shallow size distribution with $\alpha_m \approx$ 3 and 
$r_{max} \sim$ 1000 km \citep{lev2001}.  Both populations have relatively few 
objects with optical brightness $m_R \approx$ 27--27.5, which implies 
$r_0 \sim$ 20--40 km for reasonable albedo $\sim$ 0.04--0.07.  
The detached, resonant, and scattered disk populations contain large 
objects with $r_{max} \sim$ 1000 km. 
Although there are too few detached or scattered disk objects to constrain 
$\alpha_m$ or $r_0$, data for the resonant population are consistent with 
constraints derived for the hot classical population, $\alpha_m \approx$ 3 
and $r_0 \approx$ 20--40 km.

\begin{figure}[t]
\small
\centerline{\sc Table 1. Data for KBO Size Distribution}
\vskip 1.5ex
\begin{tabular}{lcccc}
\hline
\hline
KBO Class & $M_l$ ($M_{\oplus}$) & $r_{max}$ (km) & $r_0$ (km) & $q_m$ \\ 
\hline
cold cl & 0.01--0.05 & 400 & 20--40 km & $\gtrsim$ 4\\
hot cl & 0.01--0.05 & 1000 & 20--40 km & 3--3.5 \\
detached & n/a & 1500 & n/a & n/a \\
resonant & 0.01--0.05 & 1000 & 20--40 km & 3 \\
scattered & 0.1--0.3 & 700 & n/a & n/a \\ 
\hline
\end{tabular}
\end{figure}

The total mass in KBOs is a small fraction of the $\sim$ 10--30 $M_{\oplus}$ 
of solid material in a MMSN from $\sim$ 35--50 AU \citep[][see also {\em 
Chapter by Petit et al.}]{gladman:2001,bernstein:tnodist,petit:2006}.
The classical and resonant populations have $M_l \approx$ 0.01--0.1 
$M_{\oplus}$ in KBOs with $r \gtrsim$ 10--20 km. 
The scattered disk may contain more material, $M_l \sim$ 0.3 $M_{\oplus}$,
but the constraints are not as robust as for the classical and
resonant KBOs.

These data are broadly inconsistent with the predictions of self-stirring
calculations with no external perturbers.  Although self-stirring models
yield inclinations, $i \approx$ 2$^{\rm o}$--4$^{\rm o}$, close to those
observed in the cold, classical population, the small $r_{max}$ and large 
$\alpha_m$ of this group suggest that an external dynamical perturbation 
-- such as a stellar flyby or stirring by Neptune -- modified the 
evolution once $r_{max}$ reached $\sim$ 300--500 km. The observed
break radius, $r_0 \sim$ 20--40 km, also agrees better with the $r_0 \sim$ 
10 km expected from Neptune stirring calculations than the $r_0 \sim$ 1 km 
achieved in self-stirring models \citep{kb2004c,pan:2005}.  Although a 
large $r_{max}$ and small $\alpha_m$ 
for the resonant and hot, classical populations agree reasonably well with 
self-stirring models, the observed $r_{max} \sim$ 1000 km is much smaller 
than the $r_{max} \sim$ 2000--3000 km typically achieved in self-stirring 
calculations (Figure 1). Both of these populations appear to have large
$r_0$, which is also more consistent with Neptune stirring models than 
with self-stirring models.

The small $M_l$ for all populations provide additional evidence against 
self-stirring models. In the most optimistic scenario, where KBOs 
are easily broken, self-stirring models leave a factor of 5--10 more mass
in large KBOs than currently observed at 40--48 AU. Although models with 
Neptune stirring leave a factor of 2--3 more mass in KBOs at 40--48 AU
than is currently observed, Neptune ejects $\sim$ half of the KBOs at 
40--48 AU into the scattered disk \citep[e.g.,][]{duncan1995,kuchner2002}. 
With an estimated mass of 2--3 times the mass in classical and resonant 
KBOs, the scattered disk contains enough material to bridge the difference 
between the KBO mass derived from Neptune stirring models and the observed 
KBO mass.

\begin{figure}[t] \epsscale{1.0}
\plotone{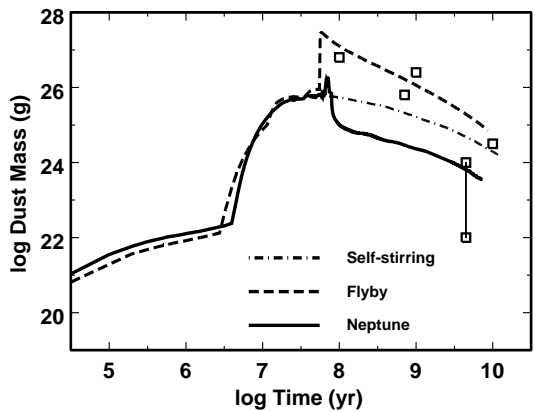}
\vskip -6ex
\caption{ \small
Evolution of mass in small dust grains (0.001--1 mm) for models with 
self-stirring (dot-dashed line), stirring from a passing star (dashed 
line), and stirring from Neptune at 30 AU (solid line) for $Q_b = 10^3$ 
erg g$^{-1}$. Calculations with smaller (large) $Q_b$ produce more (less) 
dust at $t \lesssim$ 50 Myr and somewhat more (less) dust at $t \gtrsim$ 
100 Myr.  At 1--5 Gyr, models with Neptune stirring have less dust than 
self-stirring or flyby models. The boxes show dust mass estimated for 
four nearby solar-type stars \citep[from left to right in age: HD 107146, 
$\epsilon$ Eri, $\eta$ Crv, and $\tau$ Cet;] []{greav1998,greav2004,
will2004,wyatt2005} and two estimates for the Kuiper belt \citep[boxes 
connected by solid line][]{land2002}.
}
\label{fig:mdustall}
\end{figure}

The mass in KBO dust grains provides a final piece of evidence against 
self-stirring models. From an analysis of data from Pioneer 10 and 11,
\citet{land2002} estimate a dust production rate of $\sim 10^{15}$ g yr$^{-1}$ 
in 0.01--2 mm particles at 40--50 AU. The timescale for Poynting-Robertson
drag to remove these grains from the Kuiper belt is $\sim$ 10--100 Myr
\citep{burns1979}, which yields a mass of $\sim 10^{22}$--$10^{24}$ g. 
Figure \ref{fig:mdustall} compares this dust mass with masses derived
from mid-IR and submm observations of several nearby solar-type stars
\citep{greav1998,greav2004,will2004,wyatt2005} 
and with predictions from the self-stirring, flyby, and Neptune
stirring models. The dust masses for nearby solar-type stars roughly 
follow the predictions of self-stirring models and flyby models with 
$Q_b \sim 10^3$ erg g$^{-1}$. The mass of dust in the Kuiper belt is 
1--3 orders of magnitude smaller than predicted in self-stirring models 
and is closer to the predictions of the Neptune stirring models.

To combine the dynamical properties of KBOs with these constraints,
we rely on results from $N$-body simulations that do not include 
collisional processing of small objects (see {\em Chapter} by
{\em Morbidelli et al.}). For simplicity, we consider coagulation 
in the context of the Nice model, which provides a solid framework
for interpreting the dynamics of the gas giants and the dynamical
classes of KBOs.  In the Nice model, Saturn's crossing of the 2:1 
resonance with Jupiter initiates the dynamical instability that 
populates the Kuiper belt.  As Neptune approaches $a \approx$ 30 AU, 
it captures resonant KBOs, ejects KBOs into the scattered disk and 
the Oort cloud, and excites the hot classical KBOs. Although Neptune 
might reduce the number of cold, classical KBOs formed roughly 
{\it in situ} beyond 30 AU, the properties of these KBOs probably 
reflect conditions in the Kuiper Belt when the instability began.  

The Nice model requires several results from coagulation calculations. 
Once giant planets form at 5--15 AU, collisional growth must produce 
thousands of Pluto-mass objects at 20--30 AU. Unless the planetesimal
disk was massive, growth of oligarchs must dominate collisional grinding
in this region of the disk. To produce the cold classical population at 
$\sim$ 45 AU, collisions must produce 1--10 Pluto-mass objects and then 
efficiently remove leftover planetesimals. To match the data in Table 1,
KBOs formed at 20--30 AU should have a shallower size distribution and a 
larger $r_{max}$ than those at 40--50 AU. 

Some coagulation results are consistent with the trends required in the 
Nice model. In current calculations, collisional growth naturally yields 
smaller $r_{max}$ and a steeper size distribution at larger $a$. At 
40--50 AU, Neptune-stirring models produce a few Pluto-mass objects and 
many smaller KBOs with $e \sim$ 0.1 and $i \approx$ 2$^{\rm o}$--4$^{\rm o}$. 
Although collisional growth produces more Plutos at 15--30 AU than at 
40--50 AU, collisional erosion removes material faster from the inner 
disk than from the outer disk (Fig. \ref{fig:nice}).  Thus, collisions
do not produce the thousands of Pluto-mass objects at 15--30 AU required
in the Nice model.  

Reconciling this aspect of the Nice model with the coagulation calculations 
requires a better understanding of the physical processes that can slow or 
halt the collisional cascade. Producing gas giants at 5--15 AU, thousands 
of Plutos at 20--30 AU, and a few or no Plutos at 40--50 AU 
implies that the outcome of coagulation changes markedly from 5 AU to 
50 AU. If the collisional cascade can be halted as outlined in section 
\S3.5, forming 5--10 $M_{\oplus}$ cores at 5--15 AU is straightforward.
Slowing the collisional cascade at 20--30 AU might yield a large 
population of Pluto mass objects at 20--30 AU. Because $\alpha_m$ and
$r_{max}$ are well-correlated, better constraints on the KBO size
distributions coupled with more robust coagulation calculations 
can test these aspects of the Nice model in more detail.

To conclude this section, we consider constraints on the Kuiper belt 
in the more traditional migration scenario of \citet{malhotra:1995},
where Neptune forms at $\sim$ 20--25 AU and slowly migrates to 30 AU.
To investigate the relative importance of collisional and dynamical 
depletion at 40--50 AU, \citet{charnoz:2007} couple a collision code
with a dynamical code and derive the expected distributions for size 
and orbital elements in the Kuiper belt, the scattered disk, and the
Oort cloud.
Although collisional depletion models can match the observations of
KBOs, these models are challenged to provide enough small objects into
the scattered disk and Oort cloud. Thus, the results suggest that 
dynamical mechanisms dominate collisions in removing material from 
the Kuiper belt.

Although \citet{charnoz:2007} argue against a dramatic change in collisional
evolution from 15 AU to 40 AU, the current architecture of the solar system 
provides good evidence for this possibility.  In the MMSN, the ratio of 
timescales to produce gas giant cores at 10 AU and at 25 AU is 
$\xi = (25/10)^3 \sim$ 15. In the context of the Nice model, formation 
of Saturn and Neptune at 8--11 AU in 5--10 Myr thus implies formation of 
other gas giant cores at 20--25 AU in 50--150 Myr. If these cores {\it had}
formed, they would have consumed most of the icy planetesimals at 20--30 AU,
leaving little material behind to populate the outer solar system when the
giant planets migrate. The apparent lack of gas giant core formation at 
20--30 AU indicates that the collisional cascade changed dramatically 
from 5--15 AU (where gas giant planets formed) to 20--30 AU (where gas 
giant planets did not form).  As outlined in \S3.5, understanding the 
interaction of small particles with the gas and the radiation field may
provide important insights into the evolution of oligarchic growth and thus
into the formation and structure of the solar system.

\section{\textbf{KBOs and Asteroids}}

In many ways, the Kuiper Belt is similar to the asteroid belt. Both are populations of small bodies containing relatively little mass compared to the rest of the Solar System; the structure and dynamics of both populations have been influenced significantly by the giant planets; and both have been and continue to be significantly influenced by collisions.  Due to its relative proximity to Earth, however, there are substantially more observational data available for the asteroid belt than the Kuiper Belt.  While the collisional and dynamical evolution of the asteroid belt is certainly not a solved problem, the abundance of constraints has allowed for the development of reasonably consistent models.  Here we briefly describe what is currently understood about the evolution of the asteroid belt, what insights that may give us with regards to the evolution of the Kuiper Belt, and what differences might exist in the evolution of the two populations.

It has long been recognized that the primordial asteroid belt must have contained hundreds or thousands of times more mass than the current asteroid belt \citep[e.g.][]{lecar:1973,safronov:1979,weidenschilling:1977,wetherill:1989}.  Reconstructing the initial mass distribution of the Solar System from the current masses of the planets and asteroids, for example, yields a pronounced mass deficiency in the asteroid belt region relative to an otherwise smooth distribution for the rest of the Solar System \citep{weidenschilling:1977}.  To accrete the asteroids on the timescales inferred from meteoritic evidence would require hundreds of times more mass than currently exists in the main belt \citep{wetherill:1989}. 

In addition to its pronounced mass depletion, the asteroid belt is also strongly dynamically excited.  The mean proper eccentricity and inclination of asteroids larger than $\sim$50 km in diameter are 0.135 and 10.9${}^o$ (from the catalog of \citet{knezevic:2003}), which are significantly larger than can be explained by gravitational perturbations amongst the asteroids or by simple gravitational perturbations from the planets \citep{duncan:1994}.  The fact that the different taxonomic types of asteroids (S-type, C-type, etc.) are radially mixed somewhat throughout the main belt, rather than confined to delineated zones, indicates that there has been significant scattering in semimajor axis as well \citep{gradie:1982}.

Originally, a collisional origin was suggested for the mass depletion in the asteroid belt \citep{chapman:1975}.  The difficulty of collisionally disrupting the largest asteroids, coupled with the survival of the basaltic crust of the $\sim$500-km diameter asteroid Vesta, however, suggest that collisional grinding was not the cause of the mass depletion \citep{davis:1979, davis:vesta, davis:1989, davis:collev, wetherill:1989, durda:collev1, durda:collev2, bottke:2005, bottke:fossil, obrien:numerical}.  In addition, collisional processes alone could not fully explain both the dynamical excitation and the radial mixing observed in the asteroid belt, although \citet{charnoz:2001} suggest that collisional diffusion may have contributed to its radial mixing.

Several dynamical mechanisms have been proposed to explain the mass depletion, dynamical excitation and radial mixing of the asteroid belt.  As the solar nebula dissipated, the changing gravitational potential acting on Jupiter, Saturn, and the asteroids would lead to changes in their precession rates and hence changes in the positions of secular resonances, which could `sweep' through the asteroid belt, exciting $e$ and $i$, and coupled with gas drag, could lead to semi-major axis mobility and the removal of material from the belt \citep[e.g.,][]{heppenheimer:1980,ward:1981,lemaitre:1991,lecar:1997,nagasawa:2000,nagasawa:2001,nagasawa:2002}.  It has also been suggested that sweeping secular resonances could lead to orbital excitation in the Kuiper Belt \citep{nagasawa:2000b}.  However, as reviewed by \citet{petit:2002} and \citet{obrien:2006a}, secular resonance sweeping is generally unable to simultaneously match the observed $e$ and $i$ excitation in the asteroid belt, as well as its radial mixing and mass depletion, for reasonable parameter choices (especially in the context of the Nice Model).

Another possibility is that planetary embryos were able to accrete in the asteroid belt \citep[e.g.,][]{wetherill:1992}.  The fact that Jupiter's $\sim$10 Earth-mass core was able to accrete in our Solar System beyond the asteroid belt suggests that embryos were almost certainly able to accrete in the asteroid belt, even accounting for the roughly 3-4$\times$ decrease in the mass density of solid material inside the snow line.  The scattering of asteroids by those embryos, coupled with the Jovian and Saturnian resonances in the asteroid belt, has been shown to be able to reasonably reproduce the observed $e$ and $i$ excitation in the belt as well as its radial mixing and mass depletion \citep{petit:belt_excitation2,petit:2002,obrien:2006a}.  In the majority of simulations of this scenario by both groups, the embryos are completely cleared from the asteroid belt.

Thus, the observational evidence and theoretical models for the evolution of the asteroid belt strongly suggest that dynamics, rather than collisions, dominated its mass depletion.  Collisions, however, have still played a key role in sculpting the asteroid belt.  Many dynamical families, clusterings in orbital element space, have been discovered, giving evidence for $\sim$20 breakups of 100-km or larger parent bodies over the history of the Solar System \citep{bottke:2005,bottke:fossil}.  The large 500-km diameter asteroid Vesta has a preserved basaltic crust with a single large impact basin \citep{mccord:1970, thomas:1997b}.  This basin was formed by the impact of a roughly 40-km projectile \citep{marzari:vesta, asphaug:vesta}.

\begin{figure}[t]
\plotone{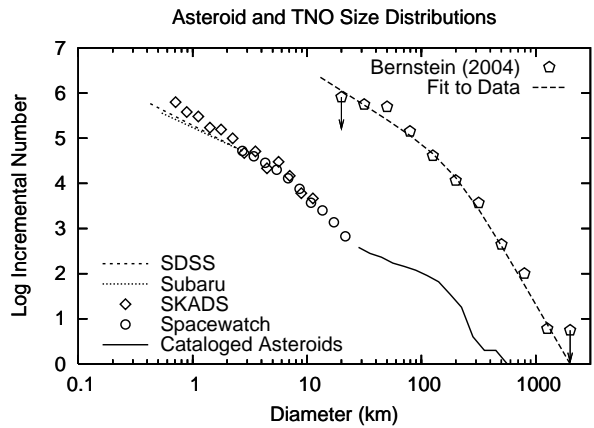}
\caption{\small Observational estimates of the main belt and TNO size distributions.  The pentagons (with dashed best-fit curve) show the total TNO population as determined from the \citet{bernstein:tnodist} HST survey, converted to approximate diameters assuming an albedo of 0.04.  Points with arrows are upper limits given by non-detections.  The solid line is the population of observed asteroids, and open circles are from debiased Spacewatch main-belt observations \citep{jedicke:mbdist}.  These data, converted to diameters, were provided by D.~Durda.  The two dashed lines are extrapolations based on the Sloan Digital Sky Survey \citep{ivezic:sdss_dist} and the Subaru Sub-km Main Belt Asteroid Survey \citep{yoshida:subaru}, and diamonds show the debiased population estimate from the SKADS survey \citep{gladman:skads}.  Error bars are left out of this plot for clarity.  Note that the TNO population is substantially more populous and massive, by roughly a factor of 1000, than the asteroid population.}
\label{fig:popplot}
\end{figure}

The size distribution of main-belt asteroids is known or reasonably constrained through observational surveys down to $\sim$1 km in diameter \citep[e.g.][]{durda:collev1,jedicke:mbdist,ivezic:sdss_dist,yoshida:subaru,gladman:skads}.  Not surprisingly, the largest uncertainties are at the smallest sizes, where good orbits are often not available for the observed asteroids, which makes the conversion to absolute magnitude and diameter difficult \citep[e.g.,][]{ivezic:sdss_dist,yoshida:subaru}.  Recent results from the Sub-Kilometer Asteroid Diameter Survey \citep[SKADS,][]{gladman:skads}, the first survey since the Palomar-Leiden Survey designed to determine orbits as well as magnitudes of main-belt asteroids, suggest that the asteroid magnitude-frequency distribution may be well represented by a single power law in the range from H=14.0 to 18.8, which corresponds to diameters of 0.7 to 7 km for an albedo of 0.11.  These observational constraints are shown in Fig.~\ref{fig:popplot} alongside the determination of the TNO size distribution from \citet{bernstein:tnodist}.

While over some size ranges, the asteroid size distribution can be fit by a single power law, over the full range of observed asteroid diameters from $\sim$1-1000 km, there are multiple bumps or kinks in the size distribution (namely around 10 and 100 km in diameter).  The change in slope of the size distribution around 100 km is due primarily to the fact that asteroids larger than this are very difficult to disrupt, and hence the size distribution of bodies larger than 100 km is likely primordial. The change in slope around 10 km has a different origin---such a structure is produced as a result of a change in the strength properties of asteroids, namely the transition from when a body's resistance to disruption is dominated by material strength to when it is dominated by self-gravity.  This transition in strength properties occurs at a size much smaller than 10 km, but results in a structure that propagates to larger sizes \citep[see, e.g.,][]{durda:collev2,obrien:analytical}.  The presence of this structure in the asteroid size distribution is consistent with the asteroids being a collisionally-relaxed population, i.e.~a population in which the size distribution has reached an approximate steady state where collisional production and collisional destruction of bodies in each size range are in balance.  

The collisional evolution of the asteroid belt has been studied by many authors \citep[e.g.][]{davis:vesta,durda:thesis,davis:collev,durda:collev1,durda:collev2,campo:collev,campo:waves,campo:2001,marzari:families}.  The most recent models of collisional evolution of the asteroid belt incorporate aspects of dynamical evolution as well, such as the removal of bodies by resonances and the Yarkovsky effect, and the enhancement in collisional activity during its massive primordial phase \citep{obrien:numerical, bottke:2005, bottke:fossil}.  In particular, \citet{bottke:2005} explicitly incorporate the results of dynamical simulations of the excitation and clearing of the main belt by embedded planetary embryos performed by \citet{petit:belt_excitation2}.  Such collisional/dynamical models can be constrained by a wide range of observational evidence such as the main belt size distribution, the number of observed asteroid families, the existence of Vesta's basaltic crust, and the cosmic ray exposure ages of ordinary chondrite meteorites, which suggest that the lifetimes of meter-scale stony bodies in the asteroid belt are on the order of 10-20 Myr \citep{marti:oc_cre_history}.

One of the most significant implications of having an early massive main belt, which was noted in early collisional models \citep[e.g.][]{chapman:1975} and recently emphasized in the case of collisional evolution plus dynamical depletion \citep[e.g.,][]{bottke:fossil}, is that the majority of the collisional evolution of the asteroid belt occurred during its early, massive phase, and there has been relatively little change in the main-belt size distribution since then.  The current, wavy main-belt size distribution, then, is a `fossil' from its first few hundred Myr of collisional and dynamical evolution.

So how does the Kuiper Belt compare to the asteroid belt in terms of its collisional and dynamical evolution?  Evidence and modeling for the asteroid belt suggest that dynamical depletion, rather than collisional erosion, was primarily responsible for reducing the mass of the primordial asteroid belt to its current level.  In the case of the Kuiper Belt, this is less clear.  As shown in Sec.~3, collisional erosion, especially when aided by stellar perturbations or the formation of Neptune, can be very effective in removing mass.  At the same time, dynamical models such as the Nice Model result in the depletion of a large amount of mass through purely dynamical means and are able to match many observational constraints.  Recent modeling that couples both collisional fragmentation and dynamical effects suggests that collisional erosion cannot play too large of a role in removing mass from the Kuiper Belt, otherwise the Scattered Disk and Oort Cloud would be too depleted to explain the observed numbers of short- and long-period comets \citep{charnoz:2007}.  That model currently does not include coagulation.  Further modeling work, which self-consistently integrates coagulation, collisional fragmentation, and dynamical effects, is necessary to fully constrain the relative contributions of collisional and dynamical depletion in the Kuiper Belt.

We have noted that the asteroid belt has a collisionally-relaxed size distribution that is not well-represented by a single power law over all size ranges.  Should we expect the same for the Kuiper Belt size distribution, and is there evidence to support this?  The collision rate in the Kuiper Belt should be roughly comparable to that in the asteroid belt, with the larger number of KBOs offsetting their lower intrinsic collision probability \citep{davis:kbo}, and as noted earlier in this chapter, the primordial Kuiper Belt, like the asteroid belt, would have been substantially more massive than the current population.  This suggests that the Kuiper Belt should have experienced a degree of collisional evolution roughly comparable to the asteroid belt, and thus is likely to be collisionally relaxed like the asteroid belt.  Observational evidence thus far is not detailed enough to say for sure if this is the case, although recent work \citep{kenyon:2004, pan:2005} suggests that the observational estimate of the TNO size distribution by \citet{bernstein:tnodist}, shown in Fig.~\ref{fig:popplot}, is consistent with a collisionally-relaxed population.

While the Kuiper Belt is likely to be collisionally relaxed, it is unlikely to mirror the exact shape of the asteroid belt size distribution.  The shape of the size distribution is determined, in part, by the strength law $Q_D^*$, which is likely to differ somewhat between asteroids and KBOs.  This is due to the difference in composition between asteroids, which are primarily rock, and KBOs, which contain a substantial amount of ice, as well as the difference in collision velocity between the two populations.  With a mean velocity of $\sim$5 km/s \citep{bottke:vel_dist}, collisions between asteroids are well into the supersonic regime (relative to the sound speed in rock).  For the Kuiper belt, collision velocities are about a factor of 5 or more smaller \citep{davis:kbo}, such that collisions between KBOs are close to the subsonic/supersonic transition.  For impacts occurring in these different velocity regimes, and into different materials, $Q_D^*$ may differ significantly \citep[e.g.,][]{benz1999}.  

The difference in collision velocity can influence the size distribution in another way as well.  With a mean collision velocity of $\sim$5 km/s, a body of a given size in the asteroid belt can collisionally disrupt a significantly larger body.  Thus, transitions in the strength properties of asteroids can lead to the formation of waves that propagate to larger sizes and manifest themselves as changes in the slope of the size distribution, as seen in Fig.~\ref{fig:popplot}.  For the Kuiper belt, with collision velocities that are about a factor of 5 or more smaller than in the asteroid belt, the difference in size between a given body and the largest body it is capable of disrupting is much smaller than in the asteroid belt, and waves should therefore be much less pronounced or non-existent in the KBO size distribution \citep[e.g.,][]{obrien:analytical}.  There is still likely to be a change in slope at the largest sizes where the population transitions from being primordial to being collisionally relaxed, and such a change appears in the debiased observational data of \citet{bernstein:tnodist} (shown here in Fig.~\ref{fig:popplot}), although recent observations suggest that the change in slope may actually occur at smaller magnitudes than found in that survey \citep{petit:2006}.

Is the size distribution of the Kuiper Belt likely to be a `fossil' like the asteroid belt?  The primordial Kuiper Belt would have been substantially more massive than the current population.  Thus, regardless of whether the depletion of its mass was primarily collisional or dynamical, collisional evolution would have been more intense early on and the majority of the collisional evolution would have occurred early in its history.  In either case, its current size distribution could then be considered a fossil from that early phase, although defining exactly when that early phase ends and the size distribution becomes `fossilized' is not equally clear in both cases.   In the case where the mass depletion of the Kuiper Belt occurs entirely through collisions, there would not necessarily be a well-defined point at which one could say that the size distribution became fossilized, as the collision rate would decay continuously with time.  In the case of dynamical depletion, where the mass would be removed fairly rapidly as in the case of the Nice Model described in Sec.~3.4, the collision rate would experience a correspondingly rapid drop, and the size distribution could be considered essentially fossilized after the dynamical depletion event.  

As noted earlier in this section, an important observable manifestation of collisions in the asteroid belt is the formation of families, i.e.~groupings of asteroids with similar orbits.  Asteroid families are thought to be the fragments of collisionally disrupted parent bodies.  These were first recognized by \citet{hirayama:1918b} who found 3 families among the 790 asteroids known at that time. The number increased to 7 families by 1926 when there were 1025 known asteroids \citep{hirayama:1927}.  Today, there are over 350,000 known asteroids while the number of asteroid families has grown to about thirty.

Given that the Kuiper Belt is likely a collisionally evolved population, are there collisional families to be found among these bodies?  Families are expected to be more difficult to recognize in the Kuiper Belt than in the asteroid belt.  Families are identified by finding statistically significant clusters of asteroid orbit elements---mainly the semi-major axis, eccentricity and inclination.  The collisional disruption of a parent bodies launches fragments with speeds of perhaps a few hundred meters/sec relative to the original target body.  This ejection speed is small compared with the orbital speeds of asteroids, hence the orbits of fragments differ by only small amounts from that of the original target body and, more importantly, from each other.  Thus, the resulting clusters of fragments are easy to identify.  

However, in the Kuiper Belt, where ejection velocities are likely to be about the same but orbital speeds are much lower, collisional disruption produces a much greater dispersion in the orbital elements of fragments.  This reduces the density of the clustering of orbital elements and makes the task of distinguishing family members from the background population much more difficult \citep{davis:kbo}.  To date, there are over 1000 KBOs known, many of which have poorly-determined orbits or are in resonances that would make the identification of a family difficult or impossible.  \citet{chiang:2003} applied lowest-order secular theory to 227 non-resonant KBOs with well-determined orbits and found no convincing evidence for a dynamical family.  Recently, however, \citet{brown2007} found evidence for a single family with at least 5 members associated with KBO 2003 EL61.  This family was identified based on the unique spectroscopic signature of its members, and confirmed by their clustered orbit elements.

Given the small numbers involved, it cannot be said whether or not finding a single KBO family at this stage is statistically that different from the original identification of 3 asteroid families when there were only 790 known asteroids \citep{hirayama:1918b}.  However, the fact that the KBO family associated with 2003 EL61 was first discovered spectroscopically, and its clustering in orbital elements was later confirmed, while nearly all asteroid families were discovered based on clusterings in orbital elements alone, suggests that even if comparable numbers of KBO families and asteroid families do exist, the greater dispersion of KBO families in orbital element space may make them more difficult to identify unless there are spectroscopic signatures connecting them as well.  

Perhaps when the number of non-resonant KBOs with good orbits approaches 1000, more populous Kuiper Belt families will be identified, and as can be done now with the asteroid belt, these KBO families can be used as constraints on the interior structures of their original parent bodies as well as on the collisional and dynamical history of the Kuiper Belt as a whole.

\section{Concluding Remarks}

Starting with a swarm of 1 m to 1 km planetesimals at 20--150 AU, the 
growth of icy planets follows a standard pattern \citep{stcol1997a, 
stcol1997b, kl1998,kl1999a,kl1999b,kb2004a,kb2004c,kb2005}.
Collisional damping and dynamical friction lead to a short period of 
runaway growth that produces 10--100 objects with $r \sim$ 300--1000 km. 
As these objects grow, they stir the orbits of leftover planetesimals up 
to the disruption velocity.  Once disruptions begin, the collisional 
cascade grinds leftover planetesimals into smaller objects faster than the 
oligarchs can accrete them. Thus, the oligarchs always contain a small 
fraction of the initial mass in solid material. For self-stirring models, 
oligarchs contain $\sim$ 10\% of the initial mass. Stellar flybys and 
stirring by a nearby gas giant augment the collisional cascade and leave 
less mass in oligarchs. The two examples in \S3.3 suggest that a very 
close flyby and stirring by Neptune leave $\sim$ 2\% to 5\% of the 
initial mass in oligarchs with $r \sim$ 100--1000 km.

This evolution differs markedly from planetary growth in the inner solar 
system. In $\sim$ 0.1--1 Myr at a few AU, runaway growth produces massive 
oligarchs, $m \gtrsim 0.01 M_{\oplus}$, that contain most of the initial 
solid mass in the disk. Aside from a few giant impacts like those that 
might produce the Moon \citep{hart1975,cam1976}, collisions remove 
little mass from these objects. 
Although the collisional cascade removes many leftover planetesimals 
before oligarchs can accrete them, the lost material is much less 
than half of the original solid mass \citep{weth1993,kb2004b}.
For $a \gtrsim$ 40 AU, runaway growth leaves most of the mass in 
0.1--10 km objects that are easily disrupted at modest collision 
velocities.  In 4.5 Gyr, the collisional cascade removes most of the 
initial disk mass inside 70--80 AU.

Together with numerical calculations of orbital dynamics ({\em Chapter} 
by {\em Morbidelli et al.}), theory now gives us a foundation for 
understanding the origin and evolution of the Kuiper belt. Within a disk 
of planetesimals at 20--100 AU, collisional growth naturally produces 
objects with $r \sim$ 10--2000 km and a size distribution reasonably close 
to that observed among KBOs. As KBOs form, migration of the giant planets 
scatters KBOs into several dynamical classes ({\em Chapter} by 
{\em Morbidelli et al.}). Once the giant planets achieve their current 
orbits, the collisional cascade reduces the total mass in KBOs to current 
levels and produces the break in the size distribution at $r \sim$ 20--40 km. 
Continued dynamical scattering by the giant planets sculpts the inner 
Kuiper belt and maintains the scattered disk.

New observations will allow us to test and to refine this theoretical 
picture. Aside from better measures of $\alpha_m$, $r_{max}$, and $r_0$
among the dynamical classes, better limits on the total mass and the 
size distribution of large KBOs with $a \sim$ 50--100 AU should yield 
a clear discriminant among theoretical models. In the Nice model, the
Kuiper belt was initially nearly empty outside of $\sim$ 50 AU. Thus,
any KBOs found with $a \sim$ 50--100 AU should have the collisional and 
dynamical signatures of the scattered disk or detached population. If 
some KBOs formed {\it in situ} at $a \gtrsim$ 50 AU, their size distribution 
depends on collisional growth modified by self-stirring and stirring by 
$\sim$ 30 $M_{\oplus}$ of large KBOs formed at 20--30 AU and scattered
through the Kuiper belt by the giant planets. From the calculations of 
Neptune stirring (\S3.3), stirring by scattered disk objects should 
yield a size distribution markedly different from the size distribution 
of detached or scattered disk objects formed at 20--30 AU. Wide-angle 
surveys on 2--3 m class telescopes (e.g., Pan-Starrs) and deep probes 
with 8--10 m telescopes can provide this test.

Information on smaller size scales -- $\alpha_d$ and $r_1$ -- place 
additional constraints on the bulk properties (fragmentation parameters)
of KBOs
and on the collisional cascade.  In any of the stirring models, there is a 
strong correlation between $r_0$, $r_1$, and the fragmentation parameters.
Thus, direct measures of $r_0$ and $r_1$ provide a clear test of KBO
formation calculations. At smaller sizes ($r \lesssim$ 0.1 km), the 
slope of the size distribution $\alpha_d$ clearly tests the fragmentation
algorithm and the ability of the collisional cascade to remove KBOs with
$r \sim$ 1--10 km.  Although the recent detection of KBOs with $r \ll$
1 km \citep{chang2006} may be an instrumental artifact 
\citep{jones2006,chang2007}, optical and X-ray occultations 
(e.g., TAOS) will eventually yield these tests. 

Finally, there is a clear need to combine coagulation and dynamical 
calculations to produce a `unified' picture of planet formation at
$a \gtrsim$ 20 AU.  \citet{charnoz:2007} provide a good start in this
direction. Because the collisional outcome is sensitive to internal 
{\it and} external dynamics, understanding the formation 
of the observed $n(r)$, $n(e)$, and $n(i)$ distributions in each 
KBO population requires treating collisional evolution and dynamics 
together.  A combined approach should yield the sensitivity of 
$\alpha_m$, $r_{max}$, and $r_0$ to the local evolution and the 
timing of the formation of giant planets, Neptune migration, and 
stellar flybys. These calculations will also test how the dynamical 
events depend on the evolution during oligarchic growth and the 
collisional cascade.  Coupled with new observations of KBOs and 
of planets and debris disks in other planetary systems, these 
calculations should give us a better understanding of the origin 
and evolution of KBOs and other objects in the outer solar system.

\vskip 6ex

We thank S. Charnoz, S. Kortenkamp, A. Morbidelli, and an anonymous
reviewer for comments that considerably improved the text. We 
acknowledge support from the NASA Astrophysics Theory Program 
(grant NAG5-13278; BCB \& SJK), the NASA Planetary Geology and 
Geophysics Program (grant NNX06AC50G; DPO), and the JPL 
Institutional Computing and Information Services and the NASA 
Directorates of Aeronautics Research, Science, Exploration Systems, 
and Space Operations (BCB \& SJK).

\bibliography{kbobib}
\bibliographystyle{natbib}

\end{document}